\documentclass[ALICE,manyauthors]{cernphprep}
\usepackage[comma,square,numbers,sort&compress]{natbib}
\usepackage{hyperref}
\usepackage{lineno}
\usepackage{xspace}
\usepackage{color}
\usepackage[T1]{fontenc}

\pdfoutput=1

\begin{document}
%

\newcommand{\pp}           {pp\xspace}
\newcommand{\ppbar}        {\mbox{$\mathrm {p\overline{p}}$}\xspace}
\newcommand{\XeXe}         {\mbox{Xe--Xe}\xspace}
\newcommand{\PbPb}         {\mbox{Pb--Pb}\xspace}
\newcommand{\pA}           {\mbox{pA}\xspace}
\newcommand{\pPb}          {\mbox{p--Pb}\xspace}
\newcommand{\AuAu}         {\mbox{Au--Au}\xspace}
\newcommand{\dAu}          {\mbox{d--Au}\xspace}

\newcommand{\s}            {\ensuremath{\sqrt{s}}\xspace}
\newcommand{\snn}          {\ensuremath{\sqrt{s_{\mathrm{NN}}}}\xspace}
\newcommand{\pt}           {\ensuremath{p_{\rm T}}\xspace}
\newcommand{\meanpt}       {$\langle p_{\mathrm{T}}\rangle$\xspace}
\newcommand{\ycms}         {\ensuremath{y_{\rm CMS}}\xspace}
\newcommand{\ylab}         {\ensuremath{y_{\rm lab}}\xspace}
\newcommand{\etarange}[1]  {\mbox{$\left | \eta \right |~<~#1$}}
\newcommand{\yrange}[1]    {\mbox{$\left | y \right |~<~#1$}}
\newcommand{\dndy}         {\ensuremath{\mathrm{d}N_\mathrm{ch}/\mathrm{d}y}\xspace}
\newcommand{\dndeta}       {\ensuremath{\mathrm{d}N_\mathrm{ch}/\mathrm{d}\eta}\xspace}
\newcommand{\avdndeta}     {\ensuremath{\langle\dndeta\rangle}\xspace}
\newcommand{\dNdy}         {\ensuremath{\mathrm{d}N_\mathrm{ch}/\mathrm{d}y}\xspace}
\newcommand{\Npart}        {\ensuremath{N_\mathrm{part}}\xspace}
\newcommand{\Ncoll}        {\ensuremath{N_\mathrm{coll}}\xspace}
\newcommand{\dEdx}         {\ensuremath{\textrm{d}E/\textrm{d}x}\xspace}
\newcommand{\RpPb}         {\ensuremath{R_{\rm pPb}}\xspace}

\newcommand{\nineH}        {$\sqrt{s}~=~0.9$~Te\kern-.1emV\xspace}
\newcommand{\seven}        {$\sqrt{s}~=~7$~Te\kern-.1emV\xspace}
\newcommand{\twoH}         {$\sqrt{s}~=~0.2$~Te\kern-.1emV\xspace}
\newcommand{\twosevensix}  {$\sqrt{s}~=~2.76$~Te\kern-.1emV\xspace}
\newcommand{\five}         {$\sqrt{s}~=~5.02$~Te\kern-.1emV\xspace}
\newcommand{\twosevensixnn}{$\sqrt{s_{\mathrm{NN}}}~=~2.76$~Te\kern-.1emV\xspace}
\newcommand{\fivenn}       {$\sqrt{s_{\mathrm{NN}}}~=~5.02$~Te\kern-.1emV\xspace}
\newcommand{\LT}           {L{\'e}vy-Tsallis\xspace}
\newcommand{\GeVc}         {Ge\kern-.1emV/$c$\xspace}
\newcommand{\MeVc}         {Me\kern-.1emV/$c$\xspace}
\newcommand{\TeV}          {Te\kern-.1emV\xspace}
\newcommand{\GeV}          {Ge\kern-.1emV\xspace}
\newcommand{\MeV}          {Me\kern-.1emV\xspace}
\newcommand{\GeVmass}      {Ge\kern-.1emV/$c^2$\xspace}
\newcommand{\MeVmass}      {Me\kern-.1emV/$c^2$\xspace}
\newcommand{\lumi}         {\ensuremath{\mathcal{L}}\xspace}

\newcommand{\ITS}          {\rm{ITS}\xspace}
\newcommand{\TOF}          {\rm{TOF}\xspace}
\newcommand{\ZDC}          {\rm{ZDC}\xspace}
\newcommand{\ZDCs}         {\rm{ZDCs}\xspace}
\newcommand{\ZNA}          {\rm{ZNA}\xspace}
\newcommand{\ZNC}          {\rm{ZNC}\xspace}
\newcommand{\SPD}          {\rm{SPD}\xspace}
\newcommand{\SDD}          {\rm{SDD}\xspace}
\newcommand{\SSD}          {\rm{SSD}\xspace}
\newcommand{\TPC}          {\rm{TPC}\xspace}
\newcommand{\TRD}          {\rm{TRD}\xspace}
\newcommand{\VZERO}        {\rm{V0}\xspace}
\newcommand{\VZEROA}       {\rm{V0A}\xspace}
\newcommand{\VZEROC}       {\rm{V0C}\xspace}
\newcommand{\Vdecay} 	   {\ensuremath{V^{0}}\xspace}

\newcommand{\ee}           {e$^+$e$^-$} 
\newcommand{\pip}          {\ensuremath{\pi^{+}}\xspace}
\newcommand{\pim}          {\ensuremath{\pi^{-}}\xspace}
\newcommand{\kap}          {\ensuremath{\rm{K}^{+}}\xspace}
\newcommand{\kam}          {\ensuremath{\rm{K}^{-}}\xspace}
\newcommand{\pbar}         {\ensuremath{\rm\overline{p}}\xspace}
\newcommand{\kzero}        {\ensuremath{{\rm K}^{0}_{\rm{S}}}\xspace}
\newcommand{\lmb}          {\ensuremath{\Lambda}\xspace}
\newcommand{\almb}         {\ensuremath{\overline{\Lambda}}\xspace}
\newcommand{\Om}           {\ensuremath{\Omega^-}\xspace}
\newcommand{\Mo}           {\ensuremath{\overline{\Omega}^+}\xspace}
\newcommand{\X}            {\ensuremath{\Xi^-}\xspace}
\newcommand{\Ix}           {\ensuremath{\overline{\Xi}^+}\xspace}
\newcommand{\Xis}          {\ensuremath{\Xi^{\pm}}\xspace}
\newcommand{\Oms}          {\ensuremath{\Omega^{\pm}}\xspace}
\newcommand{\degree}       {\ensuremath{^{\rm o}}\xspace}
\newcommand{\jpsi}         {\mbox{J/$\psi$}\xspace}

\newcommand{\flown}{\ensuremath{v_{\mathrm n}}\xspace}
\newcommand{\elflow}{\ensuremath{v_{\mathrm 2}}\xspace}
\newcommand{\triflow}{\ensuremath{v_{\mathrm 3}}\xspace}
\newcommand{\pbpb}{Pb--Pb }
\newcommand{\gevc}{GeV/$c$\xspace}
\newcommand{\raa}{\ensuremath{R_{\mathrm{AA}}}}
\newcommand{\mll}{\ensuremath{m_{\rm \ell \ell}}\xspace}
\newcommand{\mee}{\ensuremath{m_{\rm ee}}\xspace}
\newcommand{\mmumu}{\ensuremath{m_{\rm \mu \mu}}\xspace}
\newcommand{\mumu}{\ensuremath{\rm{\mu}^{+}\rm{\mu}^{-}}\xspace}

\begin{titlepage}
\PHyear{2020}       
\PHnumber{094}      
\PHdate{28 May}  

\title{\jpsi elliptic and triangular flow in Pb--Pb collisions at \fivenn}

\ShortTitle{\jpsi \elflow and \triflow in \pbpb collisions at \fivenn} 

\Collaboration{ALICE Collaboration\thanks{See Appendix~\ref{app:collab} for the list of collaboration members}}
\ShortAuthor{ALICE Collaboration} 

\begin{abstract}

The inclusive \jpsi elliptic (\elflow) and triangular (\triflow) flow coefficients measured at forward rapidity (2.5 $<y<$ 4) and the \elflow measured at midrapidity ($|y|<$ 0.9) in \pbpb collisions at \fivenn using the ALICE detector at the LHC are reported. The entire \pbpb data sample collected during Run 2 is employed, amounting to an integrated luminosity of 750 $\mu$b$^{-1}$ at forward rapidity and 93 $\mu$b$^{-1}$ at midrapidity. The results are obtained using the scalar product method and are reported as a function of transverse momentum \pt and collision centrality. At midrapidity, the \jpsi \elflow is in agreement with the forward rapidity measurement. The centrality averaged results indicate a positive \jpsi~\triflow with a significance of more than 5$\sigma$ at forward rapidity in the \pt range $2<\pt<5$~\gevc. The forward rapidity \elflow, \triflow, and \triflow/\elflow results at low and intermediate \pt ($\pt \lesssim 8$ GeV/$c$) exhibit a mass hierarchy when compared to pions and D mesons, while converging into a species-independent curve at higher \pt. At low and intermediate \pt, the results could be interpreted in terms of a later thermalization of charm quarks compared to light quarks,  while at high \pt, path-length dependent effects seem to dominate. The \jpsi \elflow measurements are further compared to a microscopic transport model calculation. Using a simplified extension of the quark scaling approach involving both light and charm quark flow components, it is shown that the D-meson \flown measurements can be described based on those for charged pions and \jpsi flow.

\end{abstract}
\end{titlepage}

\setcounter{page}{2} 

\section{Introduction} \label{section:introduction}

Ultra-relativistic heavy-ion collisions are the means to create under laboratory conditions the deconfined state of strongly-interacting matter called quark--gluon plasma (QGP). This state behaves like an ideal fluid with a shear viscosity to entropy ratio approaching the conjectured lowest possible value of $\hslash/(4\pi k_{\rm B})$  \cite{Kovtun:2004de,Voloshin:2008dg,Romatschke:2009im}. 
One of the most important observables for studying the properties of the QGP is the azimuthal dependence of particle production, also called anisotropic flow, quantified in terms of a Fourier expansion with respect to the azimuthal angle of the initial state symmetry plane for the $n$-th harmonic $\Psi_{\mathrm n}$ as
\begin{equation}
  \frac{{\mathrm d}N}{{\mathrm d} \varphi} \propto 1 + 2 \sum_{n=1}^{+ \infty} \flown\cos \left[n(\varphi-\Psi_{\mathrm n}) \right], 
  \label{eq:fourier_def_vn}
\end{equation}
where \flown is the $n$-th order harmonic coefficient and $\varphi$ is the azimuthal angle of the particles. 
The initial state spatial anisotropy of the collision overlap region is transformed into a momentum anisotropy of the produced final state particles~\cite{Ollitrault:1992bk,Voloshin:1994mz,Qiu:2011iv,Teaney:2009qa}. The medium response to the initial state anisotropy ($\varepsilon_{\mathrm n}$), which is transformed into the \flown coefficients, strongly depends on the macroscopic properties of the fireball, like the temperature dependent equation of state and the shear and bulk viscosity.

The dominant source of anisotropy is the ellipsoidal shape of the overlap region in non-central collisions that have a non-zero finite impact parameter (transverse distance separating the centers of the two nuclei), which gives rise to a large second order harmonic coefficient, \elflow, also known as elliptic flow. Fluctuations in the initial energy-density profile within the overlap region are thought to be the origin of the triangular flow, \triflow~\cite{Luzum:2008cw,Alver:2010gr,Teaney:2010vd}. Higher order harmonics are strongly damped, do not depend linearly on the initial anisotropy, and have significant contributions from the interplay of lower order harmonics~\cite{Niemi:2012aj,Gardim:2011xv,Gardim:2014tya,Acharya:2017zfg,Acharya:2020taj}.
The ALICE Collaboration published extensive studies of anisotropic flow measurements for identified light and strange particles~\cite{Abelev:2014pua,Acharya:2018zuq}. Flow coefficients for all particles show, in the low \pt range, an increasing trend with \pt mainly attributed to the radial hydrodynamic expansion of the QGP, reach a maximum in the \pt range 3--5 \gevc depending on the particle mass and species, and finally drop towards higher \pt. The behavior in the high-\pt region is commonly attributed to path-length dependent effects like energy loss~\cite{Betz:2016ayq,CMS:2012aa,Armesto:2005iq}.
At both RHIC and LHC energies, an approximate scaling of the flow coefficients  with the number of valence quarks is observed for light and strange particles~\cite{Adams:2003am,Afanasiev:2007tv,Adamczyk:2015fum,Abelev:2014pua,Acharya:2018zuq}. In the low to moderate \pt range (approximately $3 < \pt < 8$~\gevc), this scaling is hypothesized to be the consequence of the hadronization process via quark coalescence and of a common underlying partonic flow during the hydrodynamic stage of the collision~\cite{Molnar:2003ff,Lin:2003jy,Adams:2005zg,Fries:2008hs,Adams:2004bi}.

The production of charmonia, and especially of \jpsi, is one of the first proposed probes of the QGP properties, in particular the deconfinement~\cite{Matsui:1986dk}. Since charm quarks are produced during the early hard partonic collisions, they experience the entire evolution of the fireball. At the same time, their initial production cross section can be calculated in perturbative quantum chromodynamics (QCD). The suppression of the production of bound charmonium states by the free color charges of the dense deconfined medium is sensitive to both the medium bulk characteristics~\cite{Digal:2001ue,Rothkopf:2019ipj} and to the microscopic ones, like the charm-quark diffusion coefficient~\cite{Riek:2010fk,Scardina:2017ipo}.
Measurements of the \jpsi nuclear modification factor \raa~at RHIC in Au--Au collisions at $\snn=200$ GeV~\cite{Adare:2006ns} indicated a strong nuclear suppression especially for the most central collisions. At the LHC, in \pbpb collisions at $\snn=2.76$ and 5.02~TeV, the ALICE Collaboration reported a much larger \raa~compared to the one observed at RHIC~\cite{Abelev:2012rv,Abelev:2013ila,Adam:2016rdg}, despite the higher energy density present in the system. This effect is concentrated in the low-\pt region, which is consistent with charmonium regeneration by recombination of charm quarks, either at the QGP phase boundary via statistical hadronization~\cite{BraunMunzinger:2000px} or continuously throughout the fireball evolution~\cite{Du:2015wha,Du:2017qkv,Zhou:2014kka}. 

Within the statistical hadronization scenario, charm quarks thermalize in the QGP and all of the charmed bound hadrons are created at the phase boundary assuming chemical equilibration~\cite{BraunMunzinger:2000px,Andronic:2019wva}, except a small fraction created in the fireball corona that escape the medium. In transport model approaches, where charm quarks reach only a partial thermalization, roughly 50\% of the produced \jpsi originate from the recombination process, while the rest comes from primordial production~\cite{Du:2015wha,Du:2017qkv,Zhou:2014kka}. In both phenomenological approaches, it is expected that charm quarks will inherit some of the medium radial and anisotropic flow. Indeed, a significant D-meson~\cite{Sirunyan:2017plt,Acharya:2018bxo,Acharya:2017qps} and \jpsi elliptic flow~\cite{Khachatryan:2016ypw,Acharya:2017tgv,Acharya:2018pjd,Aaboud:2018ttm} was already observed at the LHC, indicating a hierarchy between the flow of charged particles, D and \jpsi mesons, with the \jpsi flow being the smallest. A positive \jpsi \elflow observed also at high \pt, typically underestimated by transport model calculations, might suggest the presence of important path length dependent effects like energy loss and the survival probability in the medium~\cite{Arleo:2017ntr,Spousta:2016agr}. In addition to \elflow, the ALICE Collaboration also published in Ref.~\cite{Acharya:2018pjd} an evidence of a positive \jpsi\triflow with a statistical significance of 3.7$\sigma$.

In this paper, the measurements of inclusive \jpsi\elflow and \triflow at forward rapidity (2.5 $<y<$ 4) and \elflow at midrapidity ($|y|<$ 0.9) in \pbpb collisions at \fivenn are discussed. Inclusive \jpsi mesons include both a prompt component from direct \jpsi production and decays of excited charmonium states and a non-prompt component from weak decays of beauty hadrons. The results are presented as a function of \pt in several collision centrality classes, expressed in percentages of the total hadronic cross section, and are compared with calculations from a microscopic transport model. The analyzed data include the full LHC Run 2 \pbpb data set, which improves the statistical precision with respect to the previous results by approximately a factor of two at forward rapidity~\cite{Acharya:2018pjd}, and a factor of nine (four) in central (semi-central) collisions at midrapidity~\cite{Acharya:2017tgv} allowing the experimental evidence of a statistically significant non-zero \jpsi\elflow at midrapidity.

\section{Experimental setup, data samples and event selection} \label{section:experimental_setups}

A detailed description of the ALICE apparatus and its performance can be found in Refs.~\cite{Aamodt:2008zz,Abelev:2014ffa}. 
At forward rapidity, \jpsi are reconstructed in the $\mumu$ decay channel with the muon spectrometer which covers the pseudorapidity range $-4 <\eta< -2.5$~\footnote{In the ALICE reference frame, the muon spectrometer covers a negative $\eta$ range and consequently a negative $y$ range. We have chosen to present our results with a positive $y$ notation, due to the symmetry of the collision system.}. 
The spectrometer includes five tracking stations, each composed of two planes of cathode pad chambers. The third station is placed inside a dipole magnet with a 3 Tm field integral.
Two trigger stations, containing two planes of resistive plate chambers each, provide single and dimuon triggers with a programmable single-muon \pt threshold.
A front absorber, made of carbon, concrete, and steel, is placed in between the primary interaction point (IP) and the first tracking station to remove primary hadrons from the collision. A second absorber, made of iron, is placed in front of the trigger chambers to further reject secondary hadrons escaping the front absorber and low-\pt muons, mainly from pion and kaon decays.
An additional conical absorber surrounds the beam pipe to protect the muon spectrometer against secondary particles produced by the interaction of large-$\eta$ particles with the beam pipe. 

At midrapidity, J/$\psi$ mesons are reconstructed in the \ee\ decay channel using the Inner Tracking System (ITS)~\cite{Aamodt:2010aa} and the Time Projection Chamber (TPC)~\cite{Alme:2010ke} in the rapidity range $|y|<0.9$. The ITS is a cylindrical-shaped  detector, consisting of 6 layers of silicon detectors used for precision tracking, reconstruction of the primary vertex of the event and event selection. The innermost two layers consists of pixels (SPD), the middle two are drift (SDD), while the two outermost layers are equipped with strip detectors (SSD). The tracklets, track segments reconstructed as pairs of hits in the SPD layers pointing to the primary vertex, are used for the determination of the event flow vector.  
The TPC is the main detector used for tracking and particle identification and consists of a cylindrical-shaped gas-filled active volume placed around the ITS. Radially, it extends between an inner radius of 0.85 m and an outer radius of 2.5 m, with a total length of 5 m along the beam axis. Particle identification in the TPC is performed via the measurement of the specific energy loss, ${\rm d}E/{\rm d}x$. 

Besides the muon spectrometer and the central barrel detectors, a set of detectors for global event characterization are also used. Two arrays of 32 scintillator counters each, covering $2.8 <\eta< 5.1$ (V0A) and $-3.7 <\eta< -1.7$ (V0C)~\cite{Abbas:2013taa}, are used for triggering, beam induced background rejection, and for the determination of the collision centrality. The 32 channels are arranged in four concentric rings with full azimuthal coverage allowing for the calculation of the event flow vector. The centrality of the events, expressed in fractions of the total inelastic hadronic cross section, is determined via a Glauber fit to the V0 amplitude as described in Refs.~\cite{centrality276,centralityNote}. In addition, two neutron Zero Degree Calorimeters~\cite{ALICE:2012aa}, installed at $\pm 112.5$ m from the nominal IP along the beam axis, are used to remove beam induced background events and electromagnetic interactions.

The analyzed data samples were collected by ALICE during the 2015 and 2018 LHC Pb--Pb runs at \fivenn using different trigger strategies for the forward muon spectrometer and the midrapidity detectors.

At forward rapidity, data were collected requiring the coincidence of the minimum bias (MB) and unlike-sign dimuon triggers. The former is defined by the coincidence of signals in the V0A and V0C arrays while the latter requires at least a pair of opposite-sign track segments in the muon trigger stations. The programmable threshold of the muon trigger algorithm was set so that the trigger efficiency for muon tracks with $\pt = 1$~\gevc is 50\% and reaches a plateau value of about 98\% at \pt $\approx$ 2.5~\gevc. 
In order to study the background, additional samples of single muon and like-sign dimuon events were also collected by requiring, in addition to the MB condition and the low-\pt threshold, at least one or a pair of same-sign track segments in the trigger system, respectively. 

At midrapidity, data were collected using the MB trigger during the 2015 data taking period, and the MB, central, and semi-central triggers in the 2018 period. The central and semi-central triggers require the MB trigger to be fired but, in addition, a condition on the total signal amplitude in the V0 detectors, corresponding to collision centralities of 0--10\% and 30--50\%, respectively, was applied. 

Both forward and midrapidity analyses require to have a primary vertex position within $\pm 10$ cm from the nominal IP along the beam axis. 
Events containing more than one collision (pile-up) are removed by exploiting the correlations between the number of clusters in the SPD, the number of reconstructed SPD tracklets, and the total signal in the V0A and V0C detectors. 
At midrapidity, events with pile-up occurring during the drift time of the TPC are rejected in the offline analysis based on the correlation between the number of SDD and SSD clusters and the total number of clusters in the TPC. The beam-induced background is filtered out offline by applying a selection based on the V0 and the ZDC timing information~\cite{Abelev:2013qoq}.

The integrated luminosity of the analyzed data samples is about 750 $\mathrm{\mu}$b$^{-1}$ for the dimuon analysis. For the measurements at midrapidity, the total luminosity recorded depends on the centrality range due to the centrality triggers, and amounts to 93 $\mathrm{\mu}$b$^{-1}$, 41 $\mathrm{\mu}$b$^{-1}$, and 20 $\mathrm{\mu}$b$^{-1}$ for the central, semi-central, and MB triggers, respectively.

\section{Data analysis} \label{section:analysis_details}

The \flown coefficients are obtained using the scalar product (SP) method~\cite{Adler:2002pu,Voloshin:2008dg}. This is a two-particle correlation technique based on the scalar product between the unit flow vector for a given harmonic $n$, ${\bf u}_{\rm n} = e^{in\varphi }$, of the particle of interest (here a dilepton) and the complex conjugate of the event flow vector in a subdetector A, ${\bf Q}_{\rm n}^{\rm A*}$. The flow coefficients are thus defined as

\begin{equation}
    v_{\rm n}\{{\rm SP}\}= \Bigg\langle {{\bf u}_{\rm n} {\bf Q}_{\rm n}^{\rm A *}} \Bigg/ \sqrt{ \frac{\langle {\bf Q}_{\rm n}^{\rm A}  {\bf Q}_{\rm n}^{\rm B *} \rangle \langle  {\bf Q}_{\rm n}^{\rm A} {\bf Q}_{\rm n}^{\rm C *} \rangle} { \langle  {\bf Q}_{\rm n}^{\rm B} {\bf Q}_{\rm n}^{\rm C *} \rangle } }   \Bigg\rangle_{\ell\ell},
    \label{eq:sp_method}
\end{equation}

where ${\bf Q}_{\rm n}^{\rm B}$ and ${\bf Q}_{\rm n}^{\rm C}$ are the $n$-th harmonic event flow vectors measured in two additional subdetectors, B and C, respectively, which are used to correct the event flow vector via the three sub-event technique~\cite{Luzum:2012da}. The star ($*$) represents the complex conjugate and the bracket $\langle ... \rangle_{\ell\ell}$ indicates the average over dileptons from all events in a given \pt range, dilepton invariant mass (\mll), and centrality interval.
The brackets $\langle ... \rangle$ in the denominator denote the average over all events in a narrow centrality interval containing the event under consideration. The V0A and V0C detectors are used in the analysis at both rapidities, while the analysis at forward rapidity uses the SPD as the third subdetector, and the analysis at midrapidity uses the TPC. As detector A, the SPD is chosen for the forward analysis and the V0C for the midrapidity one.
The V0A and V0C event flow vectors are calculated using the energy deposition measured in the individual channels. For the SPD and TPC event flow vectors, the reconstructed tracklets and the tracks are used, respectively.  

\begin{table}[h!]
\begin{center}
\caption{\label{tab:table-detectors}Summary of the details concerning the dimuon and dielectron analyses, corresponding to the forward and midrapidity region, respectively. The detectors cited in this table are described in Sec.~\ref{section:experimental_setups}, and the details concerning the three sub-event technique is presented in Sec.~\ref{section:analysis_details}.}
\begin{tabular}{cccccc}
 \hline
 \hline
 \multicolumn{2}{c}{Dilepton analysis} & \multicolumn{3}{c}{Three sub-event technique, detectors used} & Corresponding gap between \\
 \multicolumn{2}{c}{\jpsi$\rightarrow l^{+}l^{-} $}  &  A &  B &  C & ${\bf u}_{\rm n}$ and ${\bf Q}_{\rm n}^{\rm A}$ \\
 \hline
  $\mu^{+}\mu^{-}$ & $2.5 < y^{\mu\mu} < 4$ & SPD & V0A & V0C & $|\Delta \eta|>$1.1  \\
 \hline
  $\rm{e}^{+}\rm{e}^{-}$ & $|y^{\rm{ee}}| < 0.9 $ & V0C & TPC & V0A & $|\Delta \eta|>$0.8   \\
\hline
 \hline
\end{tabular}
\end{center}
\end{table}

The effects of non-uniform acceptance of the detectors used for the flow vector determination are corrected through the procedure described in Ref.~\cite{Selyuzhenkov:2007zi}. 
As was discussed in Sec.~\ref{section:experimental_setups}, the three detectors used for the event flow determination cover distinct pseudorapidity ranges, allowing for pseudorapidity gaps $\Delta \eta$ between the sub-events used for flow vector determination and dilepton reconstruction. 
The pseudorapidity gap between ${\bf u}_{\rm n}$ and ${\bf Q}_{\rm n}^{\rm A}$, corresponding to $|\Delta \eta| > 1.1$ and $|\Delta \eta| > 0.8$ for the dimuon and dielectron analysis, respectively, suppresses  the short-range correlations originating from resonance decays or jets (non-flow effects), not related to the global azimuthal anisotropy.

In the dimuon analysis, \jpsi candidates are formed by combining pairs of opposite-sign tracks reconstructed in the geometrical acceptance of the muon spectrometer using the tracking algorithm described in Ref.~\cite{Aamodt:2011gj}. The same single-muon and dimuon selection criteria used in previous analyses~\cite{Adam:2015isa,Acharya:2018pjd} are applied.
Namely, each muon track candidate should have $-4<\eta_{\rm \mu}<-2.5$, a radial transverse position at the end of the front absorber in the range $17.6<R_{\mathrm{abs}}<89.5\;\mathrm{cm}$, and  must match a track segment in the muon trigger chambers above the 1~\gevc \pt threshold. 
The rapidity of the muon pair should be within the acceptance of the muon spectrometer ($2.5 < y < 4.0$).

At midrapidity, \jpsi mesons are reconstructed in the dielectron decay channel. Electron candidates are required to be good quality tracks matched in both the ITS and the TPC, and to have a $\pt > 1$~\gevc and $|\eta|<0.9$. Tracks are selected to have at least 70 space points in the TPC, out of a maximum of 159, and a $\chi^{2}/N_{\rm dof} <$ 2 for the track fit quality. At least one hit in either of the two SPD layers is required to reject secondary electrons from photons converted in the detector material and to improve the tracking resolution. Secondary electrons are further rejected by requiring the distance-of-closest-approach (DCA) to the collision vertex to be smaller than 1~cm and 3~cm in the transverse and longitudinal directions, respectively. Electrons are identified via their specific energy loss in the TPC gas, d$E$/d$x$, by selecting a band of $\pm 3\sigma$ around the expectation value, with $\sigma$ being the d$E$/d$x$ measurement resolution. To reduce further the hadronic contamination, candidate tracks compatible within $\pm 3.5 \sigma$ with the pion or proton hypothesis are rejected.

\begin{figure}[b!]
    \begin{center}
    \includegraphics[width = 1\textwidth]{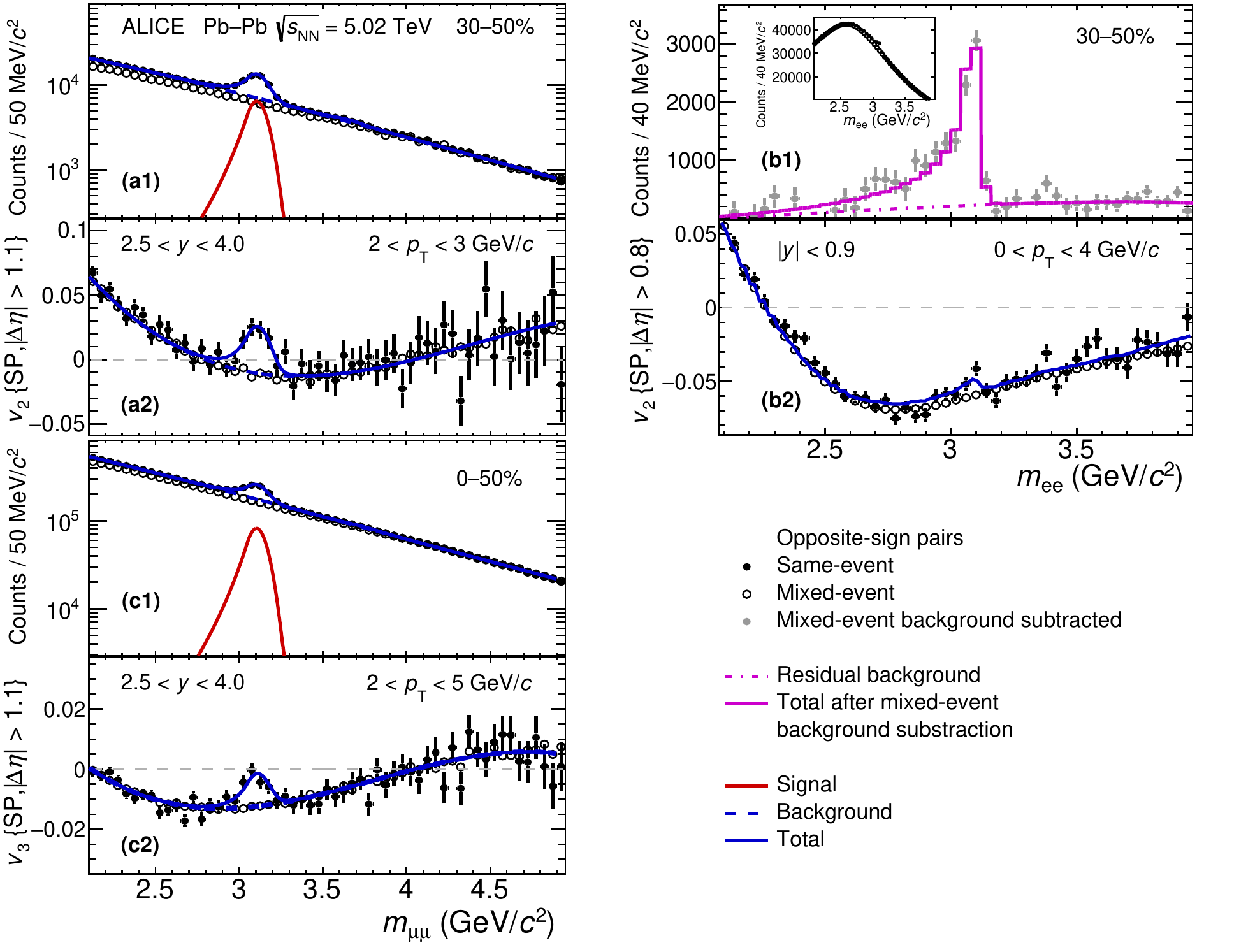}
    \end{center}
    \caption{(Color online) Invariant mass distribution (top panels (a1), (b1)) and $\elflow(\mll)$ (bottom panels (a2), (b2)) for dimuons in the ranges $2 < \pt < 3$~\gevc (top left) and for dielectrons in $0 < \pt < 4$~\gevc (top right), for the 30--50\% centrality interval. Fit functions of the invariant mass distributions and $\elflow(\mmumu)$, as discussed in Sec.~\ref{section:analysis_details}, are also shown. 
    Bottom panel, invariant mass ((c1)) and $\triflow(\mmumu)$ ((c2)) distributions for dimuons in the \pt range $2 < \pt < 5$~\gevc for the 0--50\% centrality interval. The $\flown(\mmumu)$ and $\elflow(\mee)$ distributions are plotted with the background flow obtained from the event-mixing procedure and the fit function, as discussed in the text. Only statistical uncertainties are shown.}
    \label{fig:figures_extraction}
\end{figure}

The flow coefficients are extracted from sequential fits to the dilepton invariant mass distribution, \mll, and the \flown as a function of \mll, which include the superposition of a \jpsi signal and a background contribution, using the function
\begin{equation}
    \flown(\mll) = \alpha(\mll) \hspace{0.1cm} \flown^{{\rm J}/\psi} + [1-\alpha(\mll)] \hspace{0.1cm} \flown^{\rm bkg}(\mll).
    \label{eq:vn_mass}
\end{equation}
Here, $\flown^{{\rm J}/\psi}$ denotes the \jpsi \elflow or \triflow and $\alpha(\mll)$ is the signal fraction defined as ${\rm S}/{\rm (S+B)}$. 
The latter is extracted from fits to the dilepton invariant mass distribution as described below. 
The $\flown^{{\rm bkg}}(\mll)$ corresponds to the dilepton background \elflow or \triflow.
In the dimuon analysis, the \jpsi signal is parameterized using an extended Crystal Ball (CB2) function and the background with a Variable Width Gaussian (VWG) function~\cite{ALICE-PUBLIC-2015-006}. 
In the fit, the \jpsi peak position and width are left free, while the CB2 tail parameters are fixed to the values reported in Ref.~\cite{Acharya:2017hjh}.
The signal of the $\psi(2S)$ is not included in the fit of the \flown coefficients due to its marginal significance. 
At midrapidity, the signal fraction is obtained from the dielectron invariant mass distribution in two steps. First, the combinatorial background is estimated using an event mixing technique, where pairs are built from different events with similar collision centrality, flow-vector orientation, and longitudinal position of the event vertex, and then subtracted from the same-event dielectron invariant mass distribution. The combinatorial background normalization is obtained from the ratio of the number of same-event to mixed-event like-sign pairs.  Second, the remaining distribution is fitted using a component for the signal and one for the residual background. For the \jpsi signal shape, the dielectron invariant mass distribution obtained from Monte Carlo simulations is used. The residual background, originating mainly from semileptonic decays of ${\rm c\overline{c}}$ and ${\rm b\overline{b}}$ pairs (correlated background) and imperfect matching between the same-event and mixed-event distributions, is parameterized using either a third order polynomial function at low \pt or an exponential function at high \pt.

The \flown extraction method employed in this work is described in detail in Ref.~\cite{Acharya:2018pjd}, where the $\flown^{\rm bkg}(\mll)$ distribution is obtained using an event mixing technique. 
There, it was first demonstrated that the flow coefficients of the background can be obtained from the flow coefficients of the single leptons used to form the background dileptons as 
\begin{equation}
  v_{\rm n}^{\rm bkg}(\mll) = \frac{\langle v_{\rm n}^{(1)}\cos[{\rm n}(\varphi^{(1)}-\varphi)]+v_{\rm n}^{(2)}\cos[{\rm n}(\varphi^{(2)}-\varphi)]\rangle_{\mll}}{\langle 1+2\sum\limits_{{\rm m}=1}^{\infty}{v_{\rm m}^{(1)}v_{\rm m}^{(2)}\cos[{\rm m}(\varphi^{(1)}-\varphi^{(2)}]}\rangle_{\mll}},
  \label{eq:vn_bkg}
\end{equation}
where $v_{\rm n}^{(1)}$ ($\varphi^{(1)}$) and $v_{\rm n}^{(2)}$ ($\varphi^{(2)}$) are the flow coefficients (azimuthal angles) of the two leptons, respectively, and $\varphi$ is the dilepton azimuthal angle.
The brackets $\langle \cdots \rangle_{\mll}$ denote an average over all dileptons belonging to the given $\mll$ interval.
Here, it is worth to note that the denominator in Eq.~\ref{eq:vn_bkg} represents the modification of the dilepton yields induced by the flow of single leptons.
Then, when background dileptons are built using the event mixing technique, the numerator in Eq.~\ref{eq:vn_bkg} is given by
\begin{equation}
\Big\langle  \frac{\langle {{\bf  u}_{\rm n}}^{(1)} {{\bf  Q}_{\rm n}}^{(1),{\rm A*}} \rangle }{ R_{\rm n}^{(1)}} \cos [{\rm n}(\varphi^{(1)} - \varphi)]   + \frac{\langle {{\bf  u}_{\rm n}}^{(2)} {{\bf   Q}_{\rm n}}^{(2),{\rm A*}} \rangle }{ R_{\rm n}^{(2)}} \cos [{\rm n}(\varphi^{(2)} - \varphi)]
\Big\rangle_{\mll}  . 
\label{eqn:numv2SPmethodME}
\end{equation}  
Here, ${\bf u}_{\rm n}^{(1)}$ and ${\bf u}_{\rm n}^{(2)}$ are the unit vector of the two leptons, ${\bf Q}_{\rm n}^{(1),{\rm A}}$ and ${\bf Q}_{\rm n}^{(2),{\rm A}}$ are the event flow vectors, reconstructed in detector A, of the events containing the two leptons, and $R_{\rm n}^{(1)}$ and $R_{\rm n}^{(2)}$ their respective event flow factors (corresponding to the denominator of Eq.~\ref{eq:sp_method}). 
Since the event flow vectors of the mixed events are not correlated, the mixed-event dilepton yield is not modified by the flow of the single leptons. 

Examples of fits to the invariant mass distribution (top panels corresponding to (a1), (b1), (c1)) and to $\flown(\mll)$ (bottom panels related to (a2) for $\elflow(\mmumu)$, (b2) for $\elflow(\mee)$, and (c2) for $\triflow(\mmumu)$)  are shown in Fig.~\ref{fig:figures_extraction} for the dimuon and dielectron analyses.
The background, which is mostly combinatorial, especially in central events, is well reproduced with the event mixing technique. 
In the absence of correlated background, the background flow $\flown^{\rm bkg}$ is directly given by the mixed-event flow. 
At forward rapidity, the effect of the unknown flow contribution of the correlated background and residual mismatches between the same-event and mixed-event background flow, is considered as a systematic uncertainty and is discussed in Sec.~\ref{section:systematic_uncertainties}. 
In the default approach, the flow of the correlated background is assumed to be negligible, and thus the denominator of Eq.~\ref{eq:vn_bkg} is given by the ratio $N^{\rm bkg}_{+-}/N^{\rm mix}_{+-}$ between the number of background unlike-sign dileptons $N^{\rm bkg}_{+-}$ and the number of unlike-sign dileptons from mixed events $N^{\rm mix}_{+-}$, which is obtained after a proper normalization involving like-sign dileptons as described in Ref.~\cite{Acharya:2018pjd}. 
At midrapidity, due to the smaller signal-to-background ratio, the difference between mixed and same event background flow is taken into account by considering in the fit function an additional term which accounts for the flow of the correlated background and imperfections of the mixed event procedure. This term is parameterized using a second order polynomial and acts as a correction to the background flow obtained from the mixed event procedure.


\section{Systematic uncertainties} \label{section:systematic_uncertainties}

The systematic uncertainties related to the $v_{n}$ extraction procedure, the track and event selection criteria, residual detector effects, and non-flow contributions are evaluated as described below and summarized in Tab.\ref{tab:table-syst}. A quadratic sum of the systematic uncertainties from the independent sources is used as final systematic uncertainty on the measurements.

In the dimuon analysis, the signal fraction $\alpha(\mmumu)$ is estimated by fitting the invariant mass distribution with standard signal and background functions. 
The systematic uncertainty on the determination of $\alpha(\mmumu)$ is estimated by varying the signal and background functions, as well as the mass fit range.
For the signal, in addition to a CB2, a pseudo-Gaussian with a mass-dependent width~\cite{ALICE-PUBLIC-2015-006} is also used. The tail parameters were fixed to the values obtained in Monte Carlo simulations or in other analyses with better signal significance~\cite{Acharya:2017hjh,Adam:2016rdg}. 
For the background, the VWG function was changed to a fourth order Chebyshev polynomial. The invariant mass fit range is varied from the standard $2-4$~\GeVmass to $2.6-4.6$~\GeVmass in steps of 200~\MeVmass. The corresponding systematic uncertainty for each \pt bin, evaluated as the RMS of the results of the various tests, does not exceed 0.003 for \elflow and 0.002 for \triflow. In the dielectron analysis, the fit ranges of the residual background fit are varied. No significant changes of the extracted elliptic flow are observed and no uncertainty due to the \jpsi signal extraction is assigned. 

The non-uniformity in the detector acceptance could lead to a residual effect in the calibration of the event flow vector ${\bf Q}_{\rm n}$. 
The cross-term products of the event flow vector, $\langle Q_{\rm x,A}\times Q_{\rm y,B}\rangle$, are evaluated to verify that values are negligible compared to the linear products. In addition, possible impacts on the \flown are checked by calculating the cross-term products between the components of the ${\bf Q}_{\rm n}$ vector and the unitary vector ${\bf u}_{\rm n}$ of the \jpsi candidates. 
No clear \pt or centrality dependence is found for this contribution, and the corresponding systematic uncertainty is estimated to be less than 1\%. 
Additional uncertainties related to the calculation of the reference flow vector are evaluated as the difference between the event flow factor $R_{\rm n}$ obtained using MB events or dimuon-triggered events.
For the dimuon analysis it amounts to 1\% for $R_{\rm 2}$ and up to 3\% for $R_{\rm 3}$. 

The variation of the \jpsi reconstruction efficiency with the local occupancy of the detector could bias the measured \flown. At forward rapidity, this effect is evaluated using azimuthally isotropic simulated $\jpsi \rightarrow \mumu$ decays embedded into real \pbpb events. A maximum effect of 0.002 for \elflow and 0.001 for \triflow is observed in non-central collisions with no clear \pt dependence. At midrapidity, the strongest dependence of reconstruction performance on the local detector occupancy is caused by the TPC particle identification (PID). A data driven study, using a clean electron sample from photon conversions, shows that the largest variation of the TPC electron PID response between the region along the event flow vector and the region orthogonal to it is approximately 2\% of the d$E$/d$x$ resolution. This leads to a decrease of the observed \elflow by less than 1\% and is thus neglected. 

\begin{table}[t]
\begin{center}
\caption{\label{tab:table-syst}Summary of absolute and relative (in \% of \flown) systematic uncertainties of the \jpsi \elflow and \triflow coefficients, for the dimuon and dielectron analyses. The uncertainties vary within the indicated ranges depending on the \pt bin, or centrality interval.}
\begin{tabular}{cccccc}
  \hline
  \hline
  & \multicolumn{4}{c}{$\mu^{+}\mu^{-}$}  &   $\rm{e}^{+}\rm{e}^{-}$ \\
 Sources & \elflow(\pt) & \triflow(\pt) & \elflow(Centrality) & \triflow(Centrality) &  \elflow(\pt) \\
  \hline
 Extraction method & 0--0.003 &0--0.002 & 0.001--0.004 & 0.001--0.006 &  negl  \\
 Centrality-$R_{\rm n}$ determination & 1\% & 3\% & 2\% & 3\% &  negl  \\
 Non-flow estimation & $<$1\% & negl & $<$1\% & negl &    \\
 Reconstruction efficiency & 0.001--0.002 & 0--0.001 & 0--0.002 & 0--0.001 &  negl  \\
 Correlated background shape & 0--0.009 & 0--0.015 & 0--0.010 & 0--0.011 &   \\
 TPC electron  &     &   &     &   &  0.010  \\
 identification selection &     &   &     &   &  to 0.023  \\
\hline  
\hline
\end{tabular}
\end{center}
\end{table}

The presence of a correlated background and its unknown flow contribution can affect the \flown extraction. The contribution of the correlated background to the flow of the background can be introduced in Eq.~\ref{eq:vn_bkg} by replacing the denominator $N^{\rm bkg}_{+-}/N^{\rm mix}_{+-}$ by $N^{\rm bkg}_{+-}/(N^{\rm mix}_{+-} + \beta (N^{\rm bkg}_{+-} - N^{\rm mix}_{+-}))$, where $\beta$ represents the relative strength of the correlated background flow with respect to the combinatorial background flow.
The systematic uncertainty is defined as the difference between the default fit, equivalent to $\beta = 0$, and the modified fit with $\beta$ left as a free parameter.
This uncertainty is, as expected, negligible for central collisions and low $p_{\rm T}$ but becomes significant for peripheral collisions and high $p_{\rm T}$.
The estimated systematic uncertainty for the $v_{2}$ and $v_{3}$ extraction reaches a maximum of about 0.01 for peripheral collisions and at high $p_{\rm T}$.

In the dielectron analysis, the signal-to-background ratio can vary significantly depending on the TPC electron identification selection and centrality, which may impact the \jpsi \elflow fits. Thus, the \elflow was extracted for a set of nine electron PID cuts where both the electron selection and the hadron rejection were varied such that the \jpsi efficiency is changed by approximately 50\%. The RMS of the \elflow obtained from all of these selections is assigned as a systematic uncertainty, which ranges between 0.010 and 0.023 depending on the centrality and \pt interval, while the average value is taken as central value. 
In addition, the fit range of the $\elflow(\mee)$ is varied by either making it narrower or wider, but no significant systematic effects are observed.

\section{Results and discussions}  \label{section:results}

\begin{figure}[h!]
 \centering
    \includegraphics[width = 0.95\textwidth]{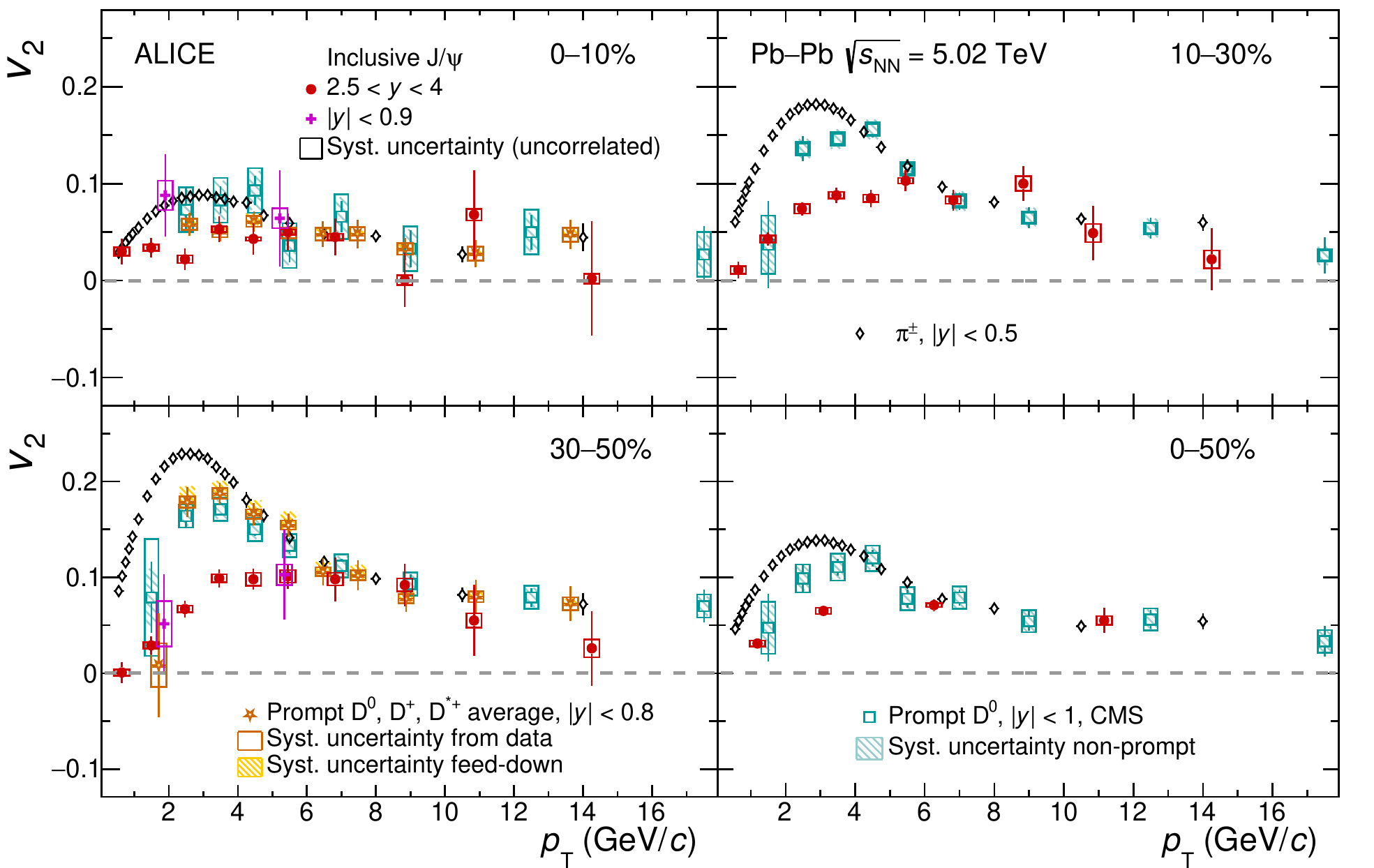}
    \caption{(Color online) Inclusive \jpsi \elflow as function of \pt in different centrality intervals (0--10\%, 10--30\%, 30--50\% and 0--50\%) in \pbpb collisions at \fivenn. Both midrapidity and forward rapidity \jpsi \elflow measurements are shown. The results are compared with the \elflow coefficients at midrapidity for charged pions~\cite{Acharya:2018zuq} and prompt ${\rm D}^{0}$ mesons~\cite{Sirunyan:2017plt,Acharya:2020zzz}. The statistical and systematic uncertainties are shown as bars and boxes, respectively. The shaded cyan boxes represent the systematic uncertainties from the contribution of non-prompt ${\rm D}^{0}$ mesons.}
    \label{fig:figure_v2vsPT_050}
\end{figure}

The \jpsi elliptic flow coefficient measured by ALICE in \pbpb collisions at \fivenn at forward and central rapidity is shown in Fig.~\ref{fig:figure_v2vsPT_050} as a function of \pt, for the centrality intervals 0--10\%, 10--30\%, 30--50\% and 0--50\%. 
Systematic uncertainties, obtained as described in the previous section, are shown as boxes around the data points, while the statistical uncertainties are shown as error bars.
Here, and in all figures as a function of $p_{\rm T}$, the \jpsi\ data points are located at the average \pt of the reconstructed \jpsi.  
These results are compared with the midrapidity \elflow measurements for charged pions by ALICE~\cite{Acharya:2018zuq} and prompt D mesons by ALICE~\cite{Acharya:2020zzz} and CMS~\cite{Sirunyan:2017plt}. At forward rapidity and for all centrality intervals, the \jpsi \elflow values increase with \pt, possibly reaching a maximum at intermediate values of \pt, and decreasing or saturating towards high \pt.
Also, the \jpsi \elflow values increase when decreasing centrality from the 0--10\% to 10--30\%, then to 30--50\%. This behavior is qualitatively similar to the one for light hadrons and D mesons. 
The \jpsi \elflow measurement at midrapidity is statistically compatible to the one at forward rapidity in both centrality intervals within uncertainties. 
Considering all the midrapidity data points as statistically independent measurements, it was found that the \jpsi \elflow is larger than zero with a significance of approximately 2.5 standard deviations in both centrality intervals.
It is worth to remark that the ALICE apparatus is undergoing an ambitious upgrade programme in preparation of Runs 3 and 4 of the LHC that will enable the separation of the prompt and non-prompt J/$\psi$ contributions to the measured flow coefficients, thus providing valuable information on both the charmonium and open beauty hadron production dynamics.

\begin{figure}[tb]
 \centering
    \includegraphics[width = 0.95\textwidth]{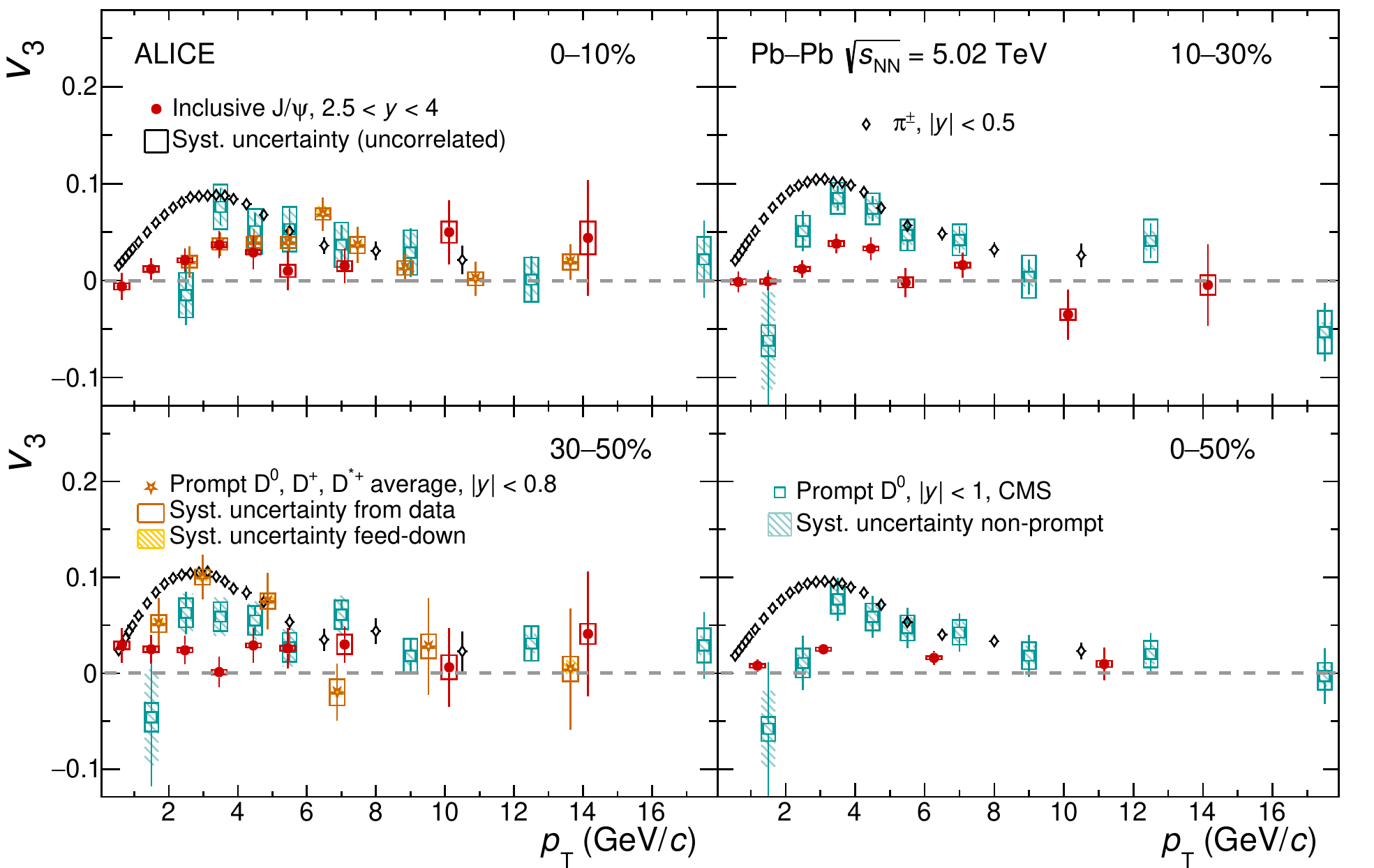}
    \caption{(Color online) Inclusive \jpsi \triflow at forward rapidity as function of \pt in different centrality intervals (0--10\%, 10--30\%, 30--50\% and 0--50\%) in \pbpb collisions at \fivenn. The results are compared to the \triflow coefficients at midrapidity for charged pions~\cite{Acharya:2018zuq} and prompt ${\rm D}^{0}$ mesons~\cite{Sirunyan:2017plt, Acharya:2020zzz}. The statistical and systematic uncertainties are shown as bars and boxes, respectively. The shaded bands represent the systematic uncertainties from the contribution of non-prompt ${\rm D}^{0}$ mesons.}
    \label{fig:figure_v3vsPT_050}
\end{figure}

As also noted previously~\cite{Acharya:2018pjd}, a clear mass hierarchy of the \elflow values is seen in the low-\pt region ($\pt<~6$~\gevc) for the light hadrons and D mesons measured at midrapidity and inclusive \jpsi, with the \jpsi exhibiting the lowest elliptic flow. Here, it is important to note that in the considered $\eta$ range, the $\eta$ dependence of the \elflow at a given \pt is expected to be negligible, as shown by the CMS measurement for charged particles~\cite{Sirunyan:2017igb}, albeit in a somewhat narrower $\eta$ range. 
At high \pt ($\pt>8 $~\gevc), the \elflow coefficients from all species converge into a single curve suggesting that, in this kinematic range, the anisotropy for all particles arises dominantly from path-length dependent energy-loss effects~\cite{Abelev:2012di}.
However, in the case of the much heavier \jpsi, one may also consider that the hydrodynamic flow, which arises from a common velocity field, still contributes significantly even at high \pt, as can be expected from the particle mass dependence of the \pt range where the flow reaches its maximum.

In Fig.~\ref{fig:figure_v3vsPT_050}, the \pt-dependent inclusive \jpsi triangular flow coefficient measured at forward rapidity is shown in each of the considered centrality intervals.
For most of the centrality and \pt intervals, the measured inclusive \jpsi \triflow is positive and with no significant centrality dependence. 
In the 0--50\% centrality range, the triangular flow coefficient is larger than zero (0.0250 $\pm$ 0.0045 (stat.) $\pm$ 0.0020 (syst.) in $2<\pt<5$ GeV/$c$) corresponding to a significance of 5.1$\sigma$, calculated adding quadratically the statistical and systematic uncertainties. 
The positive \triflow indicates that the initial state energy-density fluctuations, the dominant source of \triflow, are reflected also in the anisotropic flow of charm quarks. 
Also shown in Fig.~\ref{fig:figure_v3vsPT_050} are similar measurements for charged pions~\cite{Acharya:2018zuq} and D mesons~\cite{Sirunyan:2017plt,Acharya:2020zzz} obtained at midrapidity. 
The mass hierarchy observed for \elflow holds also in the case of \triflow. 
Together with the \jpsi \elflow, these observations provide a strong support for the hypothesis of charm quark being, at least partially, kinetically equilibrated in the dense and deconfined QGP medium.   

\begin{figure}[tb]
    \begin{center}
    \includegraphics[width = 0.95\textwidth]{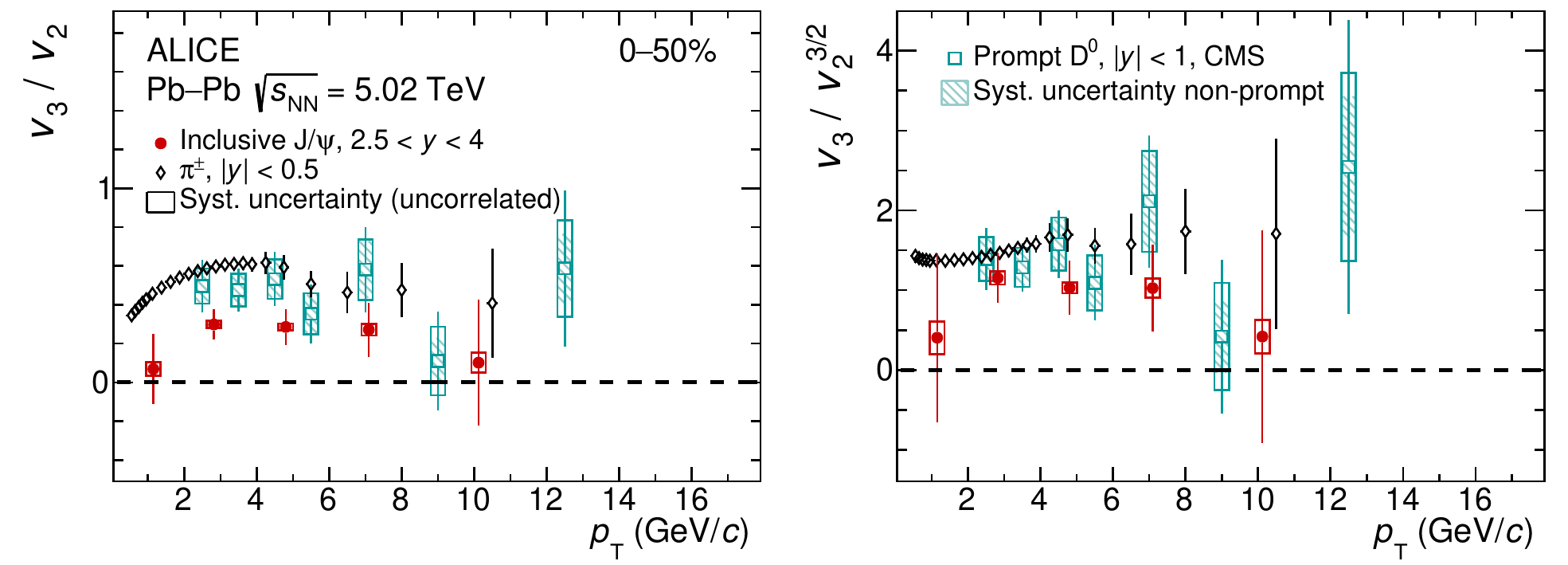}
    \end{center}
    \caption{(Color online) Ratio of \triflow to \elflow of inclusive \jpsi (left panel) and  $v_{3}/v_{2}^{3/2}$ (right panel) at forward rapidity as a function of \pt for the 0--50\% centrality interval in \pbpb collisions at \fivenn. The results are compared with the flow coefficients of charged pions~\cite{Acharya:2018zuq} and prompt ${\rm D}^{0}$ mesons at midrapidity \cite{Sirunyan:2017plt}. The statistical and systematic uncertainties are shown as bars and boxes. The shaded bands represent the systematic uncertainties from the contribution of non-prompt ${\rm D}^{0}$ mesons.}
    \label{fig:figure_v2v3RatiovsPT}
\end{figure}

The ratio of the triangular to elliptic flow coefficients, \triflow/\elflow, as a function of \pt is shown in the left panel of Fig.~\ref{fig:figure_v2v3RatiovsPT} for the inclusive \jpsi at forward rapidity, D mesons and charged pions at midrapidity. 
In this ratio, the statistical uncertainties are considered to be uncorrelated due to the weak correlation between the orientation of the ${\textbf Q}_{2}$ and ${\textbf Q}_{3}$ flow vectors~\cite{Aad:2014fla}, while the systematic uncertainties related to $\alpha(\mmumu)$ and to the reconstruction efficiency discussed in Sec.~\ref{section:systematic_uncertainties}, cancel in the ratio. The same hierarchy observed for the individual \elflow and \triflow measurements is also observed in the \triflow/\elflow ratio, which suggests that higher harmonics are damped faster for heavy quarks than for the light ones. 
At RHIC~\cite{Adams:2003zg,Adare:2010ux} and LHC~\cite{ATLAS:2012at,Acharya:2018lmh}, it was observed that the flow coefficients of light particles from different harmonics follow a power-law scaling as $v_{\rm n}^{1/\rm{n}} \propto v_{\rm m}^{1/\rm{m}}$ up to about 6~\gevc, for most centrality ranges, but the 0--5\%, independently of the harmonics n and m. The ratio $v_3/v_2^{3/2}$ in the right panel of Fig.~\ref{fig:figure_v2v3RatiovsPT} illustrates such a scaling. 
Furthermore, the $v_3/v_2^{3/2}$ for pions, D and \jpsi mesons tend to converge, although the \jpsi values are systematically lower than the ones of pions.

\begin{figure}[tb]
    \begin{center}
    \includegraphics[width = 0.6\textwidth]{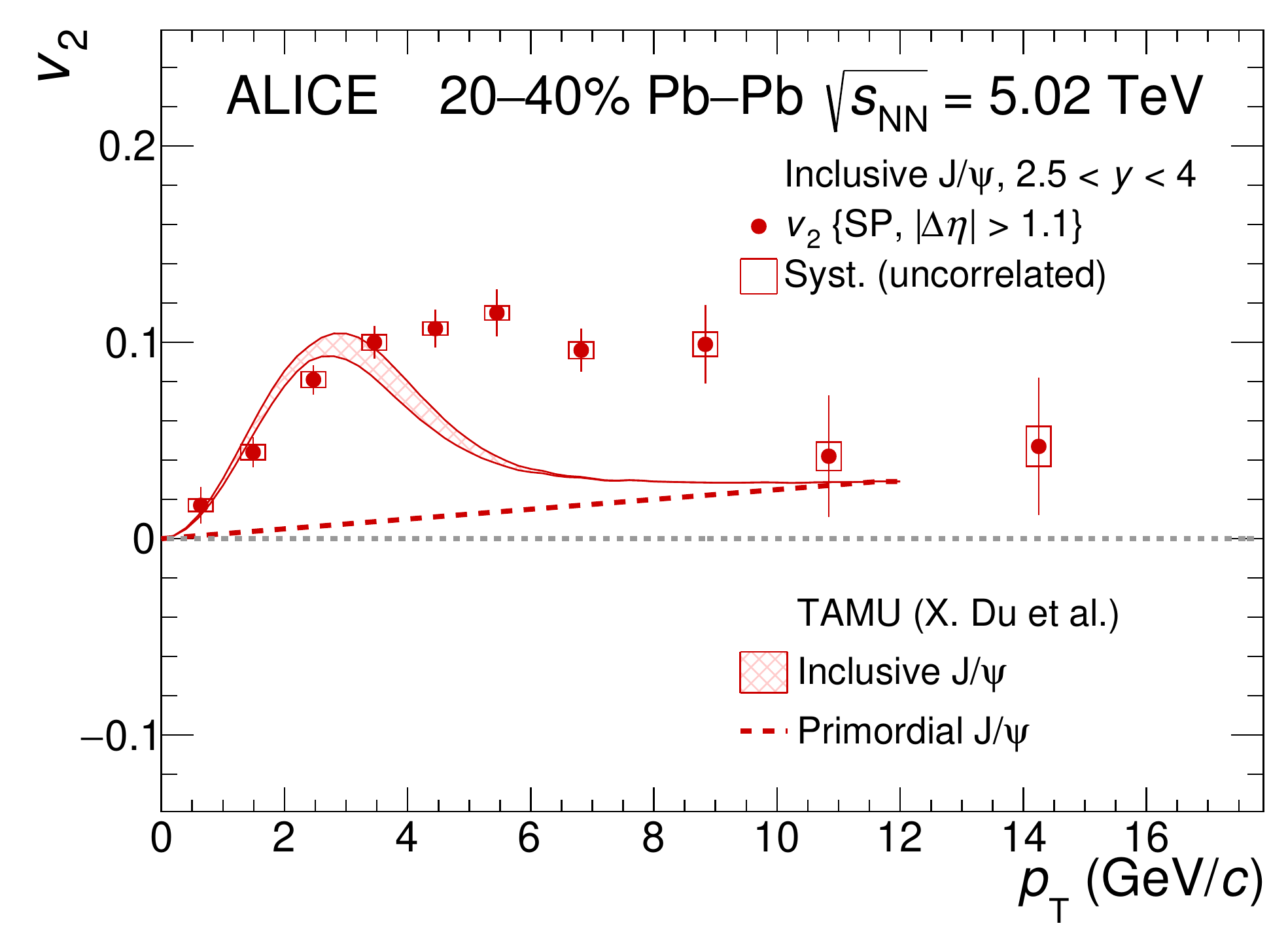}
    \end{center}
    \caption{(Color online) Inclusive J/$\psi$ $v_{2}$ as function of $p_{\rm T}$ at forward rapidity for semi-central (20--40\%) Pb--Pb collisions at \fivenn. Calculations from a transport model \cite{Du:2015wha,Du:2017qkv} are also shown.}
    \label{fig:figure_v2vsPT_models}
\end{figure}

In Fig.~\ref{fig:figure_v2vsPT_models}, the inclusive \jpsi \elflow as a function of \pt in the 20--40\% centrality interval is compared with the microscopic transport calculations by Du et al.~\cite{Du:2015wha,Du:2017qkv}.
In this model, the \jpsi are created both from the primordial hard partonic interactions but also from the recombination of thermalized charm quarks in the medium, which accounts for roughly 50\% of all \jpsi at low \pt. 
Non-prompt \jpsi mesons, created in the weak decays of beauty hadrons, are also included in the model.
The amplitude of the inclusive \jpsi \elflow in the calculations is in good agreement with the experimental measurements for \pt $<$ 4~\gevc. 
However, the overall trend of the model calculation does not describe the data well, especially in the intermediate \pt range, $4<\pt<10$~\gevc, where the \jpsi flow is largely underestimated. 
The primordial \jpsi component, which is sensitive mainly to path length dependent effects, like survival probability, exhibits a monotonically increasing trend from low towards high \pt, with this mechanism becoming the dominant source of the anisotropic flow for \pt larger than 8~\gevc. Path length dependent energy loss, widely seen as a major source of anisotropy at large \pt, is not implemented for \jpsi mesons in this calculation. It is worth noting that this model provides a qualitative good description of the centrality and transverse momentum of the \jpsi nuclear modification factor~\cite{Adam:2016rdg,Acharya:2019iur}. 

\begin{figure}[tb!]
    \begin{center}
    \includegraphics[width = 0.95\textwidth]{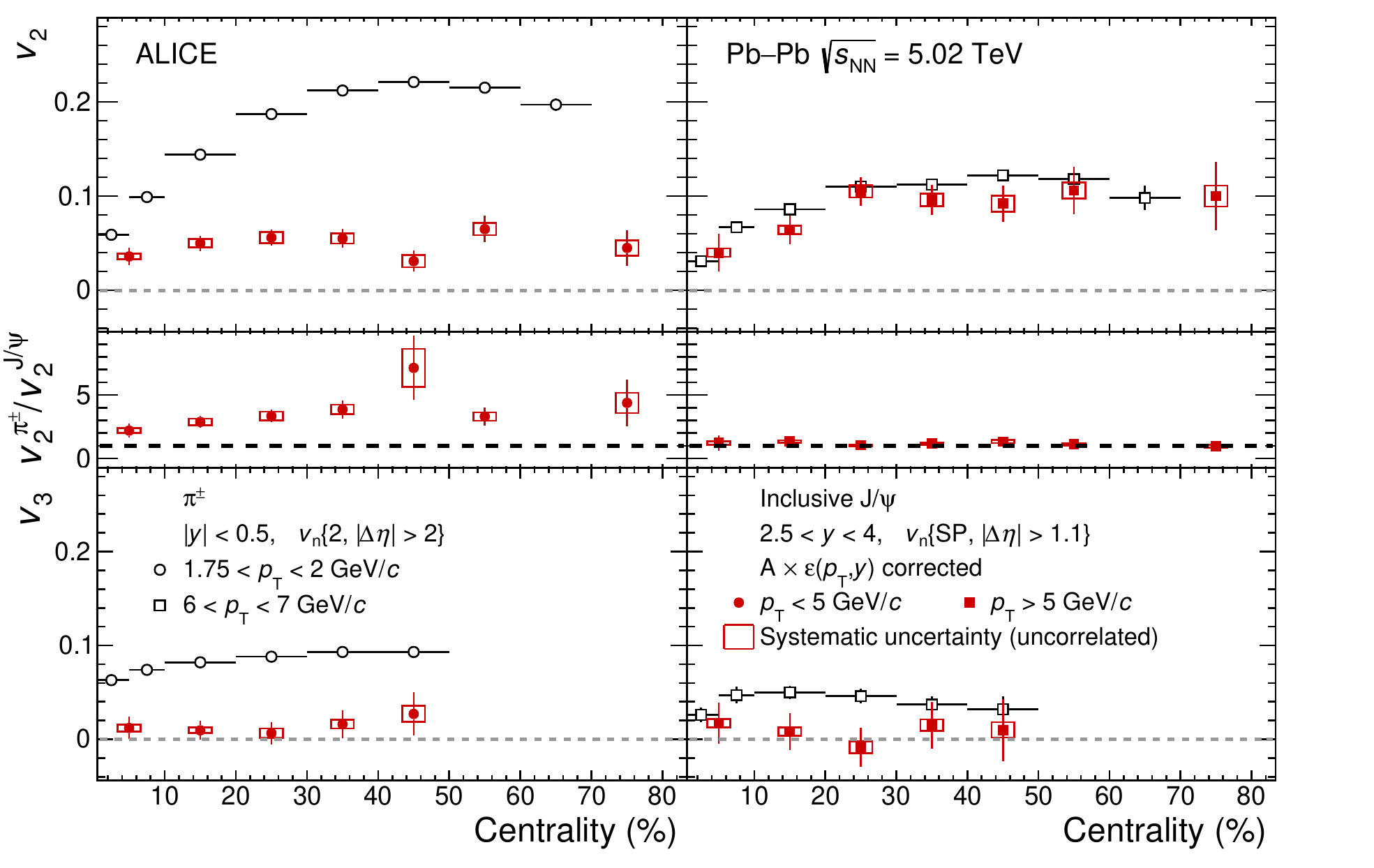}
    \end{center}
    \caption{The inclusive J/$\psi$ $v_{2}$ and $v_{3}$ as function of the centrality of the collision, at forward rapidity, for the low-$p_{\rm T}$ range $0<p_{\rm T}<5$ GeV/$c$ (left panel) and high-$p_{\rm T}$ range $5<p_{\rm T}<20$ GeV/$c$ (right panel). The results are compared to the $v_{\rm n}$ coefficients of midrapidity $\pi^{\pm}$~\cite{Acharya:2018zuq} at low and high $p_{\rm T}$ corresponding to $1.75<p_{\rm T}<2$ GeV/$c$ and $6<p_{\rm T}<7$ GeV/$c$, respectively. The ratio of midrapidity $\pi^{\pm}$ $v_{2}$ to inclusive J/$\psi$ $v_{2}$ is also shown. }
    \label{fig:figure_v2v3vsCentrality}
\end{figure}

Figure~\ref{fig:figure_v2v3vsCentrality} shows the centrality dependence of the inclusive \jpsi \elflow (top panels) and \triflow (bottom panels) for a low-\pt interval ($0 < \pt < 5$~\gevc) on the left, and a high-\pt one ($5 < \pt < 20$~\gevc) on the right. 
Here, due to the large integrated \pt range, the \flown coefficients are corrected for the \jpsi  acceptance and efficiency $A\times \varepsilon$. 
Each dimuon pair is weighted using the inverse of the \pt and $y$ dependent $A\times \varepsilon$ factor before filling the invariant mass and $v_{\rm n}(\mmumu)$ distributions.
The \pt and $y$ dependent $A\times \varepsilon$ map was obtained from the embedded simulations discussed in Section~\ref{section:systematic_uncertainties}. 
Any possible systematic uncertainty related to the $A\times \varepsilon$ corrections is already included in the systematic uncertainty due to the dependence of the reconstruction efficiency on the local detector occupancy. 
The \jpsi  results are compared with the flow coefficients of charged pions for a \pt value similar to the corrected \jpsi $\langle \pt \rangle$, published by ALICE in Ref.~\cite{Acharya:2018zuq}. In addition, the ratio $v_{\rm 2}^{{\rm \pi}}/v_{\rm 2}^{{\rm J}/\psi}$ is computed and shown in the bottom sub-panels. 
Both at low \pt ($1.75<\pt<2$~\gevc) and high \pt ($6<\pt<7$~\gevc), the \elflow of $\pi^{\pm}$ increases from central to semi-central collisions, reaching a maximum at 40--50\% centrality, and then decreases towards peripheral collisions.
For the \jpsi at low \pt, while the centrality trend is qualitatively similar, the maximum (or even saturation) of \elflow seems to be reached for more central collisions than for the pions. This is more clearly emphasized by the increasing trend of the ratio $v_{\rm 2}^{{\rm \pi}}/v_{\rm 2}^{{\rm J}/\psi}$, from central to peripheral collisions, which deviates from unity by a significance of 8.5$\sigma$. 
In the framework of transport models, this could be understood by the increasing fraction of regenerated J/$\psi$ at low $p_{\rm T}$ when moving from peripheral to central collisions.
Alternatively, and independently of the regeneration scenario, the increase of the $v_{\rm 2}^{{\rm \pi}}/v_{\rm 2}^{{\rm J}/\psi}$ from central to peripheral collisions, could also be understood in terms of partial or later thermalization of the charm quarks compared to light quarks. 
The decrease in energy density and lifetime of the system is counterbalanced by the increase of the initial spatial anisotropy towards peripheral collisions. 
The $v_{2}$ of the \jpsi will therefore reach its maximum at more central collisions compared to light particles because charm quarks require larger energy densities to develop flow~\cite{Scardina:2017ipo,Rapp:2018qla,Song:2015sfa,Cao:2011et}.
At high \pt, \jpsi mesons and charged pions seem to exhibit the same centrality dependence, although the $v_{\rm 2}$ coefficients are systematically lower for the \jpsi mesons than for the pions. 
Such a similar centrality dependence could indicate a similar mechanism at the origin of the flow for both \jpsi mesons and pions at high \pt.

The centrality dependence of the \triflow coefficient at low \pt is less pronounced than that of the \elflow for both pions and \jpsi, as expected since initial state fluctuations only weakly depend on centrality. Also, the \jpsi \triflow is smaller relative to the one of charged pions, in both the \pt intervals considered.

\begin{figure}[tb]
    \begin{center}
    \includegraphics[width = 0.95\textwidth]{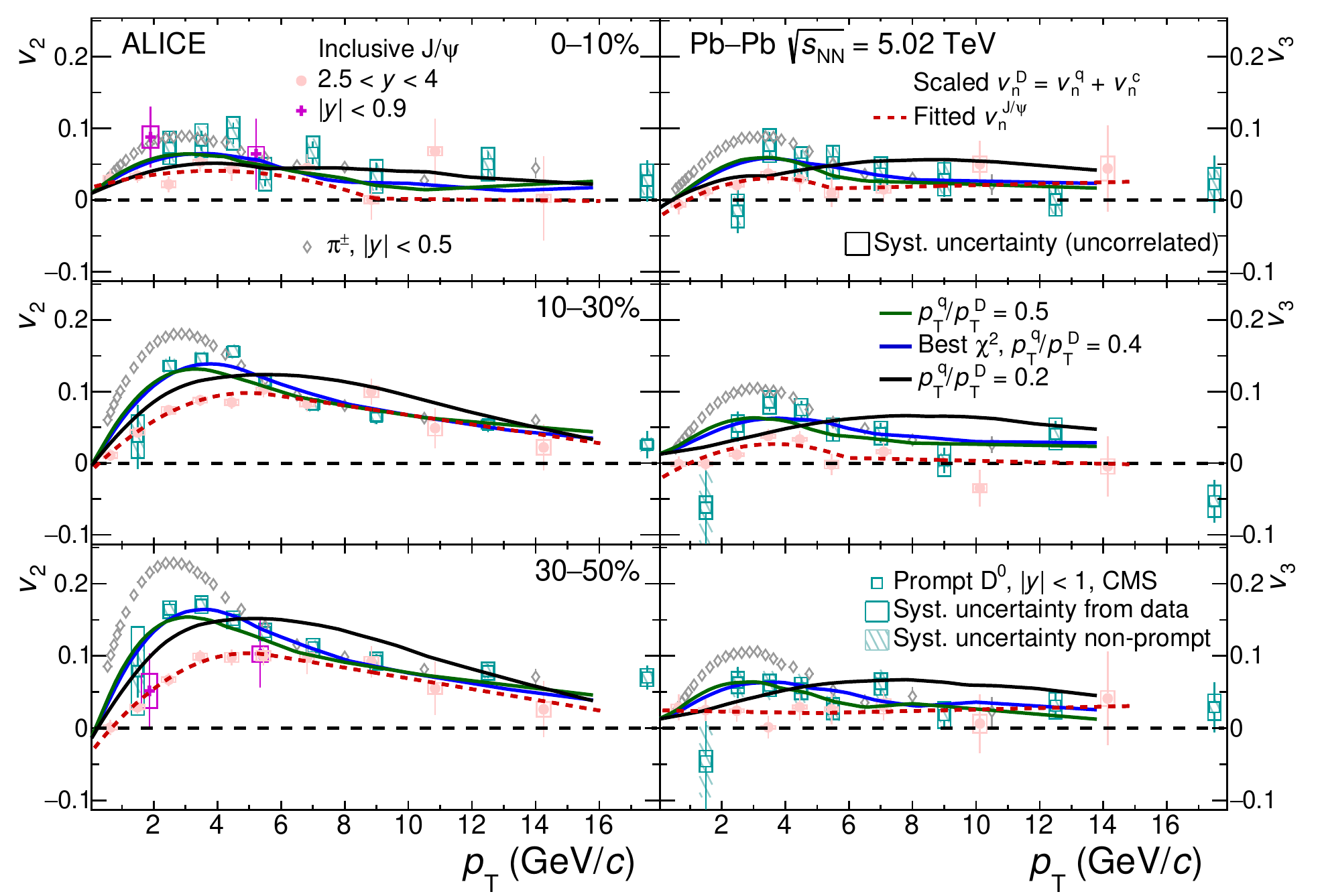}
    \end{center}
    \caption{(Color online) Elliptic (left panels) and triangular (right panels) flow of inclusive \jpsi , D mesons~\cite{Sirunyan:2017plt} and charged pions as a function of \pt for the centrality intervals 0--10\% (top), 10--30\% (middle) and 30--50\% (bottom). The continuous curves show the calculated D-meson flow based on different values of the \pt fraction carried by the light quark (see text). The red dashed curves show the fits to the \jpsi \flown using ad-hoc functions (see text).}
    \label{fig:figure_scalingCharmFlow}
\end{figure}

The flow of light and strange particles was shown to approximately scale with the number of constituent quarks (NCQ scaling) at both RHIC and LHC energies \cite{Zheng:2016iia,Singha:2016aim}. This was typically interpreted to arise naturally in hadronization scenarios based on quark coalescence in which the flow of bound mesons and baryons depends solely on the collective flow of light and strange quarks (assumed to be identical) and the number of valence quarks \cite{Molnar:2003ff,Adams:2004bi}. In the case of charmed hadrons, the NCQ scaling assuming a flavor independent flow would obviously not work due to the large observed differences between the flow of light-flavor particles, D and \jpsi mesons. However, this scaling can be extended by assuming that the much heavier charm quark has a different flow magnitude \cite{Lin:2003jy} and that it can be derived from the flow of the \jpsi via the usual NCQ formula, $v_{\rm n}^{\rm J/\psi}(\pt^{\rm J/\psi}) = 2 \cdot v_{\rm n}^{\rm c}(\pt^{\rm J/\psi}/2)$. Then it is straightforward to show that the flow of the D meson can be constructed as the sum of the flow coefficients for light and charm quarks as
\begin{equation}
v_{\rm n}^{\rm D}(p_{\rm T}^{\rm D}) = v_{\rm n}^{\rm q}(p_{\rm T}^{\rm q}) + v_{\rm n}^{\rm c}(p_{\rm T}^{\rm c}),
\label{eq:quarkScaling}
\end{equation}
where $p_{\rm T}^{\rm q}$ and $p_{\rm T}^{\rm c}$ are the \pt of the light and charm quarks, respectively, corresponding to the D-meson \pt, $p_{\rm T}^{\rm D}$. The light quark flow is obtained by interpolating the measured charged pions flow using $v_{\rm n}^{\rm \pi}(\pt^{\rm \pi}) = 2 \cdot v_{\rm n}^{\rm q}(\pt^{\rm \pi}/2)$.
Figure~\ref{fig:figure_scalingCharmFlow} shows a comparison of the D-meson \elflow and \triflow as a function of \pt, derived assuming the above described procedure, to the D-meson \flown measured by the CMS Collaboration~\cite{Sirunyan:2017plt}. 

The red dashed curves show fits to the \jpsi \flown employing an ad-hoc function, a third order polynomial at low \pt and a linear function at high \pt, used to extract the flow of charm quarks needed to obtain the scaled D-meson flow according to Eq.~\ref{eq:quarkScaling}. 
The scaled D-meson flow is found to be very sensitive to the fraction of \pt carried by each of the constituent quarks. In coalescence-like models, constituent quarks must have equal velocities which leads to a sharing of the D-meson \pt proportional to the effective quark masses. This implies that by far the largest fraction of \pt should be carried by the charm quark. Based on the simplistic and naive approach described here, a \pt sharing between light and charm quarks~\cite{Jia:2006vj,Lin:2003jy} where the ratio $p_{\rm T}^{\rm q}/p_{\rm T}^{\rm D} = 0.2$ (black curve), is clearly disfavored by the data. Surprisingly, it was found that a good description of the D-meson flow measurement, as illustrated by the blue curves in Fig.~\ref{fig:figure_scalingCharmFlow}, is obtained when the light quark carries a relatively large fraction of the D-meson \pt (dark blue and green curves). 
The best agreement with the D-meson data of the CMS Collaboration~\cite{Sirunyan:2017plt} is obtained when the light-quark \pt fraction has a value of $p_{\rm T}^{\rm q}/p_{\rm T}^{\rm D}$ = 0.4 (dark blue curve), but a rather good description of the data is observed also when assuming that the light and charm quarks share equally the D-meson \pt (green curve). 
Within uncertainties, the scaling seems to work well for both \elflow and \triflow over the entire covered \pt range and in all centrality intervals.

\section{Conclusion}

In summary, the inclusive \jpsi \elflow at forward and midrapidity and the \jpsi \triflow at forward rapidity were measured in \pbpb collisions at \fivenn using the scalar product method.
In non-central collisions, the \jpsi \elflow values are found to be positive up to the last interval corresponding to $12<\pt<20$~\gevc and reach a maximum of approximately 0.1 around a \pt of 5~\gevc.
The \jpsi \triflow values at forward rapidity reach 0.04 around a \pt of 4~\gevc and are positive in the 0--50\% centrality interval for $2<\pt<5$~\gevc with a significance of 5.1$\sigma$. 
The mass hierarchy observed for \elflow, $v_{\rm 2,\pi}>v_{\rm 2,D}>v_{\rm 2,J/\psi}$, seems to also hold in the case of \triflow and will be the subject of more detailed studies with the Run 3 and Run 4 data.
At high \pt, the \elflow for all particles converge to similar values, suggesting that path-length dependent effects become dominant there. 
The measured \jpsi \triflow/\elflow ratios exhibits the same hierarchy indicating that higher harmonics are damped faster for charmonia compared to lighter particles.
The \pt-integrated $v_2$ coefficient in a low and a high-\pt region is in both cases dependent on centrality and reaches a maximum value of about 0.1, while the $v_3$ has no clear centrality dependence. 
Both J/$\psi$ \pt-integrated $v_2$ and $v_3$ coefficients, either at low $p_{\rm T}$ or at high $p_{\rm T}$ are found to be lower than the ones of charged pions at a \pt similar to the \jpsi average \pt. 
At low \pt, the ratio of the charged pions $v_2$ to those of \pt-integrated J/$\psi$ increase from central to peripheral collisions, compatible with a scenario in which charm quarks thermalize later than the light ones. At high \pt, this ratio is compatible with unity without any statistically significant centrality dependence.

Using an extension of the well known number of constituent quark scaling, the measured charged pion and J/$\psi$ $v_{\rm n}$ can be used as proxies in order to derive the D-meson \elflow and \triflow as a combination of the flow of light and charm quarks. Within this procedure, it is surprising to observe that the measured D-meson \elflow and \triflow can be described if one considers that the light and charm quarks share similar fractions of the D-meson $p_{\rm T}$, which is counterintuitive in a coalescence approach. 
The fact that such a simple scaling works suggests that the flow of charmonia and open charm mesons can be effectively explained assuming a common underlying charm quark flow in addition to the flow of light quarks.

\newenvironment{acknowledgement}{\relax}{\relax}

\begin{acknowledgement}
\section*{Acknowledgements}

The ALICE Collaboration would like to thank all its engineers and technicians for their invaluable contributions to the construction of the experiment and the CERN accelerator teams for the outstanding performance of the LHC complex.
The ALICE Collaboration gratefully acknowledges the resources and support provided by all Grid centres and the Worldwide LHC Computing Grid (WLCG) collaboration.
The ALICE Collaboration acknowledges the following funding agencies for their support in building and running the ALICE detector:
A. I. Alikhanyan National Science Laboratory (Yerevan Physics Institute) Foundation (ANSL), State Committee of Science and World Federation of Scientists (WFS), Armenia;
Austrian Academy of Sciences, Austrian Science Fund (FWF): [M 2467-N36] and Nationalstiftung f\"{u}r Forschung, Technologie und Entwicklung, Austria;
Ministry of Communications and High Technologies, National Nuclear Research Center, Azerbaijan;
Conselho Nacional de Desenvolvimento Cient\'{\i}fico e Tecnol\'{o}gico (CNPq), Financiadora de Estudos e Projetos (Finep), Funda\c{c}\~{a}o de Amparo \`{a} Pesquisa do Estado de S\~{a}o Paulo (FAPESP) and Universidade Federal do Rio Grande do Sul (UFRGS), Brazil;
Ministry of Education of China (MOEC) , Ministry of Science \& Technology of China (MSTC) and National Natural Science Foundation of China (NSFC), China;
Ministry of Science and Education and Croatian Science Foundation, Croatia;
Centro de Aplicaciones Tecnol\'{o}gicas y Desarrollo Nuclear (CEADEN), Cubaenerg\'{\i}a, Cuba;
Ministry of Education, Youth and Sports of the Czech Republic, Czech Republic;
The Danish Council for Independent Research | Natural Sciences, the VILLUM FONDEN and Danish National Research Foundation (DNRF), Denmark;
Helsinki Institute of Physics (HIP), Finland;
Commissariat \`{a} l'Energie Atomique (CEA) and Institut National de Physique Nucl\'{e}aire et de Physique des Particules (IN2P3) and Centre National de la Recherche Scientifique (CNRS), France;
Bundesministerium f\"{u}r Bildung und Forschung (BMBF) and GSI Helmholtzzentrum f\"{u}r Schwerionenforschung GmbH, Germany;
General Secretariat for Research and Technology, Ministry of Education, Research and Religions, Greece;
National Research, Development and Innovation Office, Hungary;
Department of Atomic Energy Government of India (DAE), Department of Science and Technology, Government of India (DST), University Grants Commission, Government of India (UGC) and Council of Scientific and Industrial Research (CSIR), India;
Indonesian Institute of Science, Indonesia;
Centro Fermi - Museo Storico della Fisica e Centro Studi e Ricerche Enrico Fermi and Istituto Nazionale di Fisica Nucleare (INFN), Italy;
Institute for Innovative Science and Technology , Nagasaki Institute of Applied Science (IIST), Japanese Ministry of Education, Culture, Sports, Science and Technology (MEXT) and Japan Society for the Promotion of Science (JSPS) KAKENHI, Japan;
Consejo Nacional de Ciencia (CONACYT) y Tecnolog\'{i}a, through Fondo de Cooperaci\'{o}n Internacional en Ciencia y Tecnolog\'{i}a (FONCICYT) and Direcci\'{o}n General de Asuntos del Personal Academico (DGAPA), Mexico;
Nederlandse Organisatie voor Wetenschappelijk Onderzoek (NWO), Netherlands;
The Research Council of Norway, Norway;
Commission on Science and Technology for Sustainable Development in the South (COMSATS), Pakistan;
Pontificia Universidad Cat\'{o}lica del Per\'{u}, Peru;
Ministry of Science and Higher Education, National Science Centre and WUT ID-UB, Poland;
Korea Institute of Science and Technology Information and National Research Foundation of Korea (NRF), Republic of Korea;
Ministry of Education and Scientific Research, Institute of Atomic Physics and Ministry of Research and Innovation and Institute of Atomic Physics, Romania;
Joint Institute for Nuclear Research (JINR), Ministry of Education and Science of the Russian Federation, National Research Centre Kurchatov Institute, Russian Science Foundation and Russian Foundation for Basic Research, Russia;
Ministry of Education, Science, Research and Sport of the Slovak Republic, Slovakia;
National Research Foundation of South Africa, South Africa;
Swedish Research Council (VR) and Knut \& Alice Wallenberg Foundation (KAW), Sweden;
European Organization for Nuclear Research, Switzerland;
Suranaree University of Technology (SUT), National Science and Technology Development Agency (NSDTA) and Office of the Higher Education Commission under NRU project of Thailand, Thailand;
Turkish Atomic Energy Agency (TAEK), Turkey;
National Academy of  Sciences of Ukraine, Ukraine;
Science and Technology Facilities Council (STFC), United Kingdom;
National Science Foundation of the United States of America (NSF) and United States Department of Energy, Office of Nuclear Physics (DOE NP), United States of America.
\end{acknowledgement}

\bibliographystyle{utphys}   
\bibliography{bibliography}

\newpage
\appendix

%
%

\section{The ALICE Collaboration}
\label{app:collab}

\begingroup
\small
\begin{flushleft}
S.~Acharya\Irefn{org141}\And 
D.~Adamov\'{a}\Irefn{org95}\And 
A.~Adler\Irefn{org74}\And 
J.~Adolfsson\Irefn{org81}\And 
M.M.~Aggarwal\Irefn{org100}\And 
G.~Aglieri Rinella\Irefn{org34}\And 
M.~Agnello\Irefn{org30}\And 
N.~Agrawal\Irefn{org10}\textsuperscript{,}\Irefn{org54}\And 
Z.~Ahammed\Irefn{org141}\And 
S.~Ahmad\Irefn{org16}
% ...
\And
\renewcommand\labelenumi{\textsuperscript{\theenumi}~}

\section*{Affiliation notes}
\renewcommand\theenumi{\roman{enumi}}
\begin{Authlist}
\item \Adef{org*}Deceased
\item \Adef{orgI}Italian National Agency for New Technologies, Energy and Sustainable Economic Development (ENEA), Bologna, Italy
\item \Adef{orgII}Dipartimento DET del Politecnico di Torino, Turin, Italy
\item \Adef{orgIII}M.V. Lomonosov Moscow State University, D.V. Skobeltsyn Institute of Nuclear, Physics, Moscow, Russia
\item \Adef{orgIV}Department of Applied Physics, Aligarh Muslim University, Aligarh, India
\item \Adef{orgV}Institute of Theoretical Physics, University of Wroclaw, Poland
\end{Authlist}

\section*{Collaboration Institutes}
\renewcommand\theenumi{\arabic{enumi}~}
\begin{Authlist}
\item \Idef{org1}A.I. Alikhanyan National Science Laboratory (Yerevan Physics Institute) Foundation, Yerevan, Armenia
\item \Idef{org2}Bogolyubov Institute for Theoretical Physics, National Academy of Sciences of Ukraine, Kiev, Ukraine
\item \Idef{org3}Bose Institute, Department of Physics  and Centre for Astroparticle Physics and Space Science (CAPSS), Kolkata, India
\item \Idef{org4}Budker Institute for Nuclear Physics, Novosibirsk, Russia
\item \Idef{org5}California Polytechnic State University, San Luis Obispo, California, United States
\item \Idef{org6}Central China Normal University, Wuhan, China
\item \Idef{org7}Centre de Calcul de l'IN2P3, Villeurbanne, Lyon, France
\item \Idef{org8}Centro de Aplicaciones Tecnol\'{o}gicas y Desarrollo Nuclear (CEADEN), Havana, Cuba
\item \Idef{org9}Centro de Investigaci\'{o}n y de Estudios Avanzados (CINVESTAV), Mexico City and M\'{e}rida, Mexico
\item \Idef{org10}Centro Fermi - Museo Storico della Fisica e Centro Studi e Ricerche ``Enrico Fermi', Rome, Italy
\item \Idef{org11}Chicago State University, Chicago, Illinois, United States
\item \Idef{org12}China Institute of Atomic Energy, Beijing, China
\item \Idef{org13}Comenius University Bratislava, Faculty of Mathematics, Physics and Informatics, Bratislava, Slovakia
\item \Idef{org14}COMSATS University Islamabad, Islamabad, Pakistan
\item \Idef{org15}Creighton University, Omaha, Nebraska, United States
\item \Idef{org16}Department of Physics, Aligarh Muslim University, Aligarh, India
\item \Idef{org17}Department of Physics, Pusan National University, Pusan, Republic of Korea
\item \Idef{org18}Department of Physics, Sejong University, Seoul, Republic of Korea
\item \Idef{org19}Department of Physics, University of California, Berkeley, California, United States
\item \Idef{org20}Department of Physics, University of Oslo, Oslo, Norway
\item \Idef{org21}Department of Physics and Technology, University of Bergen, Bergen, Norway
\item \Idef{org22}Dipartimento di Fisica dell'Universit\`{a} 'La Sapienza' and Sezione INFN, Rome, Italy
\item \Idef{org23}Dipartimento di Fisica dell'Universit\`{a} and Sezione INFN, Cagliari, Italy
\item \Idef{org24}Dipartimento di Fisica dell'Universit\`{a} and Sezione INFN, Trieste, Italy
\item \Idef{org25}Dipartimento di Fisica dell'Universit\`{a} and Sezione INFN, Turin, Italy
\item \Idef{org26}Dipartimento di Fisica e Astronomia dell'Universit\`{a} and Sezione INFN, Bologna, Italy
\item \Idef{org27}Dipartimento di Fisica e Astronomia dell'Universit\`{a} and Sezione INFN, Catania, Italy
\item \Idef{org28}Dipartimento di Fisica e Astronomia dell'Universit\`{a} and Sezione INFN, Padova, Italy
\item \Idef{org29}Dipartimento di Fisica `E.R.~Caianiello' dell'Universit\`{a} and Gruppo Collegato INFN, Salerno, Italy
\item \Idef{org30}Dipartimento DISAT del Politecnico and Sezione INFN, Turin, Italy
\item \Idef{org31}Dipartimento di Scienze e Innovazione Tecnologica dell'Universit\`{a} del Piemonte Orientale and INFN Sezione di Torino, Alessandria, Italy
\item \Idef{org32}Dipartimento di Scienze MIFT, Universit\`{a} di Messina, Messina, Italy
\item \Idef{org33}Dipartimento Interateneo di Fisica `M.~Merlin' and Sezione INFN, Bari, Italy
\item \Idef{org34}European Organization for Nuclear Research (CERN), Geneva, Switzerland
\item \Idef{org35}Faculty of Electrical Engineering, Mechanical Engineering and Naval Architecture, University of Split, Split, Croatia
\item \Idef{org36}Faculty of Engineering and Science, Western Norway University of Applied Sciences, Bergen, Norway
\item \Idef{org37}Faculty of Nuclear Sciences and Physical Engineering, Czech Technical University in Prague, Prague, Czech Republic
\item \Idef{org38}Faculty of Science, P.J.~\v{S}af\'{a}rik University, Ko\v{s}ice, Slovakia
\item \Idef{org39}Frankfurt Institute for Advanced Studies, Johann Wolfgang Goethe-Universit\"{a}t Frankfurt, Frankfurt, Germany
\item \Idef{org40}Fudan University, Shanghai, China
\item \Idef{org41}Gangneung-Wonju National University, Gangneung, Republic of Korea
\item \Idef{org42}Gauhati University, Department of Physics, Guwahati, India
\item \Idef{org43}Helmholtz-Institut f\"{u}r Strahlen- und Kernphysik, Rheinische Friedrich-Wilhelms-Universit\"{a}t Bonn, Bonn, Germany
\item \Idef{org44}Helsinki Institute of Physics (HIP), Helsinki, Finland
\item \Idef{org45}High Energy Physics Group,  Universidad Aut\'{o}noma de Puebla, Puebla, Mexico
\item \Idef{org46}Hiroshima University, Hiroshima, Japan
\item \Idef{org47}Hochschule Worms, Zentrum  f\"{u}r Technologietransfer und Telekommunikation (ZTT), Worms, Germany
\item \Idef{org48}Horia Hulubei National Institute of Physics and Nuclear Engineering, Bucharest, Romania
\item \Idef{org49}Indian Institute of Technology Bombay (IIT), Mumbai, India
\item \Idef{org50}Indian Institute of Technology Indore, Indore, India
\item \Idef{org51}Indonesian Institute of Sciences, Jakarta, Indonesia
\item \Idef{org52}INFN, Laboratori Nazionali di Frascati, Frascati, Italy
\item \Idef{org53}INFN, Sezione di Bari, Bari, Italy
\item \Idef{org54}INFN, Sezione di Bologna, Bologna, Italy
\item \Idef{org55}INFN, Sezione di Cagliari, Cagliari, Italy
\item \Idef{org56}INFN, Sezione di Catania, Catania, Italy
\item \Idef{org57}INFN, Sezione di Padova, Padova, Italy
\item \Idef{org58}INFN, Sezione di Roma, Rome, Italy
\item \Idef{org59}INFN, Sezione di Torino, Turin, Italy
\item \Idef{org60}INFN, Sezione di Trieste, Trieste, Italy
\item \Idef{org61}Inha University, Incheon, Republic of Korea
\item \Idef{org62}Institute for Nuclear Research, Academy of Sciences, Moscow, Russia
\item \Idef{org63}Institute for Subatomic Physics, Utrecht University/Nikhef, Utrecht, Netherlands
\item \Idef{org64}Institute of Experimental Physics, Slovak Academy of Sciences, Ko\v{s}ice, Slovakia
\item \Idef{org65}Institute of Physics, Homi Bhabha National Institute, Bhubaneswar, India
\item \Idef{org66}Institute of Physics of the Czech Academy of Sciences, Prague, Czech Republic
\item \Idef{org67}Institute of Space Science (ISS), Bucharest, Romania
\item \Idef{org68}Institut f\"{u}r Kernphysik, Johann Wolfgang Goethe-Universit\"{a}t Frankfurt, Frankfurt, Germany
\item \Idef{org69}Instituto de Ciencias Nucleares, Universidad Nacional Aut\'{o}noma de M\'{e}xico, Mexico City, Mexico
\item \Idef{org70}Instituto de F\'{i}sica, Universidade Federal do Rio Grande do Sul (UFRGS), Porto Alegre, Brazil
\item \Idef{org71}Instituto de F\'{\i}sica, Universidad Nacional Aut\'{o}noma de M\'{e}xico, Mexico City, Mexico
\item \Idef{org72}iThemba LABS, National Research Foundation, Somerset West, South Africa
\item \Idef{org73}Jeonbuk National University, Jeonju, Republic of Korea
\item \Idef{org74}Johann-Wolfgang-Goethe Universit\"{a}t Frankfurt Institut f\"{u}r Informatik, Fachbereich Informatik und Mathematik, Frankfurt, Germany
\item \Idef{org75}Joint Institute for Nuclear Research (JINR), Dubna, Russia
\item \Idef{org76}Korea Institute of Science and Technology Information, Daejeon, Republic of Korea
\item \Idef{org77}KTO Karatay University, Konya, Turkey
\item \Idef{org78}Laboratoire de Physique des 2 Infinis, Ir\`{e}ne Joliot-Curie, Orsay, France
\item \Idef{org79}Laboratoire de Physique Subatomique et de Cosmologie, Universit\'{e} Grenoble-Alpes, CNRS-IN2P3, Grenoble, France
\item \Idef{org80}Lawrence Berkeley National Laboratory, Berkeley, California, United States
\item \Idef{org81}Lund University Department of Physics, Division of Particle Physics, Lund, Sweden
\item \Idef{org82}Nagasaki Institute of Applied Science, Nagasaki, Japan
\item \Idef{org83}Nara Women{'}s University (NWU), Nara, Japan
\item \Idef{org84}National and Kapodistrian University of Athens, School of Science, Department of Physics , Athens, Greece
\item \Idef{org85}National Centre for Nuclear Research, Warsaw, Poland
\item \Idef{org86}National Institute of Science Education and Research, Homi Bhabha National Institute, Jatni, India
\item \Idef{org87}National Nuclear Research Center, Baku, Azerbaijan
\item \Idef{org88}National Research Centre Kurchatov Institute, Moscow, Russia
\item \Idef{org89}Niels Bohr Institute, University of Copenhagen, Copenhagen, Denmark
\item \Idef{org90}Nikhef, National institute for subatomic physics, Amsterdam, Netherlands
\item \Idef{org91}NRC Kurchatov Institute IHEP, Protvino, Russia
\item \Idef{org92}NRC \guillemotleft Kurchatov\guillemotright~Institute - ITEP, Moscow, Russia
\item \Idef{org93}NRNU Moscow Engineering Physics Institute, Moscow, Russia
\item \Idef{org94}Nuclear Physics Group, STFC Daresbury Laboratory, Daresbury, United Kingdom
\item \Idef{org95}Nuclear Physics Institute of the Czech Academy of Sciences, \v{R}e\v{z} u Prahy, Czech Republic
\item \Idef{org96}Oak Ridge National Laboratory, Oak Ridge, Tennessee, United States
\item \Idef{org97}Ohio State University, Columbus, Ohio, United States
\item \Idef{org98}Petersburg Nuclear Physics Institute, Gatchina, Russia
\item \Idef{org99}Physics department, Faculty of science, University of Zagreb, Zagreb, Croatia
\item \Idef{org100}Physics Department, Panjab University, Chandigarh, India
\item \Idef{org101}Physics Department, University of Jammu, Jammu, India
\item \Idef{org102}Physics Department, University of Rajasthan, Jaipur, India
\item \Idef{org103}Physikalisches Institut, Eberhard-Karls-Universit\"{a}t T\"{u}bingen, T\"{u}bingen, Germany
\item \Idef{org104}Physikalisches Institut, Ruprecht-Karls-Universit\"{a}t Heidelberg, Heidelberg, Germany
\item \Idef{org105}Physik Department, Technische Universit\"{a}t M\"{u}nchen, Munich, Germany
\item \Idef{org106}Politecnico di Bari, Bari, Italy
\item \Idef{org107}Research Division and ExtreMe Matter Institute EMMI, GSI Helmholtzzentrum f\"ur Schwerionenforschung GmbH, Darmstadt, Germany
\item \Idef{org108}Rudjer Bo\v{s}kovi\'{c} Institute, Zagreb, Croatia
\item \Idef{org109}Russian Federal Nuclear Center (VNIIEF), Sarov, Russia
\item \Idef{org110}Saha Institute of Nuclear Physics, Homi Bhabha National Institute, Kolkata, India
\item \Idef{org111}School of Physics and Astronomy, University of Birmingham, Birmingham, United Kingdom
\item \Idef{org112}Secci\'{o}n F\'{\i}sica, Departamento de Ciencias, Pontificia Universidad Cat\'{o}lica del Per\'{u}, Lima, Peru
\item \Idef{org113}St. Petersburg State University, St. Petersburg, Russia
\item \Idef{org114}Stefan Meyer Institut f\"{u}r Subatomare Physik (SMI), Vienna, Austria
\item \Idef{org115}SUBATECH, IMT Atlantique, Universit\'{e} de Nantes, CNRS-IN2P3, Nantes, France
\item \Idef{org116}Suranaree University of Technology, Nakhon Ratchasima, Thailand
\item \Idef{org117}Technical University of Ko\v{s}ice, Ko\v{s}ice, Slovakia
\item \Idef{org118}The Henryk Niewodniczanski Institute of Nuclear Physics, Polish Academy of Sciences, Cracow, Poland
\item \Idef{org119}The University of Texas at Austin, Austin, Texas, United States
\item \Idef{org120}Universidad Aut\'{o}noma de Sinaloa, Culiac\'{a}n, Mexico
\item \Idef{org121}Universidade de S\~{a}o Paulo (USP), S\~{a}o Paulo, Brazil
\item \Idef{org122}Universidade Estadual de Campinas (UNICAMP), Campinas, Brazil
\item \Idef{org123}Universidade Federal do ABC, Santo Andre, Brazil
\item \Idef{org124}University of Cape Town, Cape Town, South Africa
\item \Idef{org125}University of Houston, Houston, Texas, United States
\item \Idef{org126}University of Jyv\"{a}skyl\"{a}, Jyv\"{a}skyl\"{a}, Finland
\item \Idef{org127}University of Liverpool, Liverpool, United Kingdom
\item \Idef{org128}University of Science and Technology of China, Hefei, China
\item \Idef{org129}University of South-Eastern Norway, Tonsberg, Norway
\item \Idef{org130}University of Tennessee, Knoxville, Tennessee, United States
\item \Idef{org131}University of the Witwatersrand, Johannesburg, South Africa
\item \Idef{org132}University of Tokyo, Tokyo, Japan
\item \Idef{org133}University of Tsukuba, Tsukuba, Japan
\item \Idef{org134}Universit\'{e} Clermont Auvergne, CNRS/IN2P3, LPC, Clermont-Ferrand, France
\item \Idef{org135}Universit\'{e} de Lyon, Universit\'{e} Lyon 1, CNRS/IN2P3, IPN-Lyon, Villeurbanne, Lyon, France
\item \Idef{org136}Universit\'{e} de Strasbourg, CNRS, IPHC UMR 7178, F-67000 Strasbourg, France, Strasbourg, France
\item \Idef{org137}Universit\'{e} Paris-Saclay Centre d'Etudes de Saclay (CEA), IRFU, D\'{e}partment de Physique Nucl\'{e}aire (DPhN), Saclay, France
\item \Idef{org138}Universit\`{a} degli Studi di Foggia, Foggia, Italy
\item \Idef{org139}Universit\`{a} degli Studi di Pavia, Pavia, Italy
\item \Idef{org140}Universit\`{a} di Brescia, Brescia, Italy
\item \Idef{org141}Variable Energy Cyclotron Centre, Homi Bhabha National Institute, Kolkata, India
\item \Idef{org142}Warsaw University of Technology, Warsaw, Poland
\item \Idef{org143}Wayne State University, Detroit, Michigan, United States
\item \Idef{org144}Westf\"{a}lische Wilhelms-Universit\"{a}t M\"{u}nster, Institut f\"{u}r Kernphysik, M\"{u}nster, Germany
\item \Idef{org145}Wigner Research Centre for Physics, Budapest, Hungary
\item \Idef{org146}Yale University, New Haven, Connecticut, United States
\item \Idef{org147}Yonsei University, Seoul, Republic of Korea
\end{Authlist}
\endgroup
  
\end{document}